\documentclass{article}[12pt]
\usepackage[a4paper, margin=1in]{geometry}

\usepackage{url}
\usepackage{hyperref}
\hypersetup{
    colorlinks=true,
    linkcolor=blue,
    filecolor=blue,      
    urlcolor =blue,
    citecolor=blue}
\usepackage{xcolor}
\usepackage{float}
\usepackage{amsmath}
\usepackage{amssymb}
\usepackage{algorithm}
\usepackage{algpseudocodex}
\usepackage{graphicx}
\usepackage{booktabs}
\usepackage{amsthm}
\usepackage{tikz}
\usepackage[capitalize, nameinlink]{cleveref}
\usepackage{authblk}
\usepackage{cancel}

\newtheorem{theorem}{Theorem}[section]

\newtheorem{definition}[theorem]{Definition}

\usepackage[square,sort,comma,super,authoryear]{natbib}
\usepackage{accents} 

\newcommand{\D}{\mathrm{d}}

\newcommand{\tL}{{\widetilde{L}}}

\newcommand{\tR}{{\widetilde{R}}}

\newcommand{\tV}{{\widetilde{V}}}

\newcommand{\tmu}{{\widetilde{\mu}}}

\newcommand{\tsigma}{{\widetilde{\sigma}}}

\newcommand{\ba}{{\boldsymbol{a}}}

\newcommand{\bb}{{\boldsymbol{b}}}

\newcommand{\bg}{{\boldsymbol{g}}}

\newcommand{\bK}{{\boldsymbol{K}}}

\newcommand{\bM}{{\boldsymbol{M}}}

\newcommand{\bu}{{\boldsymbol{u}}}

\newcommand{\bv}{{\boldsymbol{v}}}

\newcommand{\bx}{{\boldsymbol{x}}}
\newcommand{\bX}{{\boldsymbol{X}}}

\newcommand{\bY}{{\boldsymbol{Y}}}
\newcommand{\bz}{{\boldsymbol{z}}}

\newcommand{\bbeta}{{\boldsymbol{\beta}}}

\newcommand{\bDelta}{{\boldsymbol{\Delta}}}

\newcommand{\bEta}{{\boldsymbol{\eta}}}
\newcommand{\btheta}{{\boldsymbol{\theta}}}

\newcommand{\blambda}{{\boldsymbol{\lambda}}}

\newcommand{\tba}{{\widetilde{\boldsymbol{a}}}}

\newcommand{\hn}{{\widehat{n}}}

\newcommand{\hV}{{\widehat{V}}}

\newcommand{\hmu}{{\widehat{\mu}}}
\newcommand{\hnu}{{\widehat{\nu}}}

\newcommand{\hsigma}{{\widehat{\sigma}}}

\newcommand{\ma}{{\mathsf{a}}}
\newcommand{\mA}{{\mathsf{A}}}
\newcommand{\mb}{{\mathsf{b}}}

\newcommand{\mE}{{\mathsf{E}}}

\newcommand{\mG}{{\mathsf{G}}}

\newcommand{\mi}{{\mathsf{i}}}
\newcommand{\mI}{{\mathsf{I}}}

\newcommand{\mK}{{\mathsf{K}}}

\newcommand{\mx}{{\mathsf{x}}}
\newcommand{\mX}{{\mathsf{X}}}

\newcommand{\mLambda}{{\mathsf{\Lambda}}}

\newcommand{\mSigma}{{\mathsf{\Sigma}}}

\newcommand{\calK}{{\mathcal{K}}}

\newcommand{\calL}{{\mathcal{L}}}

\newcommand{\calN}{{\mathcal{N}}}

\newcommand{\calO}{{\mathcal{O}}}

\newcommand{\calU}{{\mathcal{U}}}

\newcommand{\bbC}{{\mathbb{C}}}

\newcommand{\bbE}{{\mathbb{E}}}

\newcommand{\bbN}{{\mathbb{N}}}

\newcommand{\bbR}{{\mathbb{R}}}

\newcommand{\bbV}{{\mathbb{V}}}

\newcommand{\bzero}{{\boldsymbol{0}}}
\newcommand{\bone}{{\boldsymbol{1}}}

\newcommand{\simiid}{\overset{\mathrm{IID}}{\sim}}

\newcommand{\Cov}{\mathrm{Cov}}

\newcommand{\qtq}[1]{\quad\mathrm{#1}\quad}
\newcommand{\qqtqq}[1]{\qquad\mathrm{#1}\qquad}

\DeclareMathOperator*{\argmax}{argmax}

\newcommand{\starhat}[1]{\accentset{\star}{#1}}

\newcommand\ringring[1]{%
  {
   \mathop{\kern0pt #1}\limits^{
     \vbox to-1.85ex{
       \kern-2ex 
       \hbox to 0pt{\hss\normalfont\kern.1em \r{}\kern-.45em \r{}\hss}%
       \vss 
     }
   }
  }
}

\usepackage[textsize=tiny]{todonotes}

\title{Fast Bayesian Multilevel Quasi-Monte Carlo}
\author[1]{Aleksei G. Sorokin}
\author[2]{Pieterjan Robbe}
\author[2]{Gianluca Geraci}
\author[2]{\\ Michael S. Eldred}
\author[1]{Fred J. Hickernell}
\affil[1]{Illinois Institute of Technology, Department of Applied Mathematics, United States}
\affil[2]{Sandia National Laboratories}
\date{}

\begin{document}


\newpage 

\maketitle

\begin{abstract}
    Existing multilevel quasi-Monte Carlo (MLQMC) methods often rely on multiple independent randomizations of a low-discrepancy (LD) sequence to estimate statistical errors on each level. While this approach is standard, it can be less efficient than simply increasing the number of points from a single LD sequence. However, a single LD sequence does not permit statistical error estimates in the current framework. We propose to recast the MLQMC problem in a Bayesian cubature framework, which uses a single LD sequence and quantifies numerical error through the posterior variance of a Gaussian process (GP) model. When paired with certain LD sequences, GP regression and hyperparameter optimization can be carried out at only $\mathcal{O}(n \log n)$ cost, where $n$ is the number of samples. Building on the adaptive sample allocation used in traditional MLQMC, where the number of samples is doubled on the level with the greatest expected benefit, we introduce a new Bayesian utility function that balances the computational cost of doubling against the anticipated reduction in posterior uncertainty. We also propose a new digitally-shift-invariant (DSI) kernel of adaptive smoothness, which combines multiple higher-order DSI kernels through a weighted sum of smoothness parameters, for use with fast digital net GPs. A series of numerical experiments illustrate the performance of our fast Bayesian MLQMC method and error estimates for both single-level problems and multilevel problems with a fixed number of levels. The Bayesian error estimates obtained using digital nets are found to be reliable, although, in some cases, mildly conservative.
\end{abstract}

\section{Introduction}

Monte Carlo methods are invaluable tools for approximating the expectation of a simulation subject to high-dimensional random sources. Standard Monte Carlo methods sample the simulation at independent and identically distributed (IID) points and compute a sample mean estimate with $\calO(n^{-1/2})$ error in the number of points $n$. Quasi-Monte Carlo (QMC) replaces IID points with highly uniform, low-discrepancy (LD) sequences which enable convergence like $\calO(n^{-1+\delta})$ or better where $\delta>0$ is arbitrarily small \citep{hickernell.qmc_what_why_how,dick.high_dim_integration_qmc_way,owen.mc_book,owen.mc_book_practical,niederreiter.qmc_book,caflisch1998monte,dick.digital_nets_sequences_book,dick2022lattice}. \Cref{fig:points} compares IID points against extensible LD sequences which have highly uniform coverage and fill in gaps left by previous samples as the number of points increases through powers of two.

Multilevel Monte Carlo (MLMC) methods exploit cheaper low-fidelity simulations to accelerate approximation of the maximum-fidelity expectation \citep{giles.MLMC_path_simulation,giles2015multilevel}. Specifically, the expectation at the maximum level is often written as the sum of expectations of differences between consecutive levels, each of which is approximated using a Monte Carlo method. Multilevel quasi-Monte Carlo (MLQMC) \citep{giles.mlqmc_path_simulation,robbe2019multilevel} methods approximate each of these expectations using QMC. They are typically deployed in algorithms which iteratively double the number of LD points on the level of maximum utility as defined through the cost of sampling on each level and the estimated error on each level. Such MLQMC algorithms require QMC algorithms which provide both accurate approximations and reliable error estimates. 

Perhaps the most common method of QMC error estimation is to take $R$ independent randomizations of an LD sequence and construct statistical error estimates based on the resulting $R$ independent sample means \citep{owen.mc_book_practical}. Recently, \citet{lecuyer.RQMC_CLT_bootstrap_comparison} ran a comprehensive series of numerical experiments which showcased the reliability and robustness of this RQMC method when using $R \geq 10$ randomizations. For a fixed budget, the reliability of the RQMC error estimator increases with $R$, but the accuracy of the underlying approximation decreases with $R$ as fewer LD points must be randomized more times. In other words, for RQMC, the number of randomizations $R$ trades off approximation accuracy for error estimation reliability.

One method to overcome the need for multiple randomizations in QMC error estimation is to cast the problem into a Bayesian cubature framework \citep{briol2019probabilistic,o1991bayes,rasmussen2003bayesian,briol.frank_wolfe_bayesian_quadrature}. Here, the integrand is assumed to be a Gaussian process (GP), so the QMC error may be quantified in terms of the GP's posterior variance. \citet{rathinavel.bayesian_QMC_lattice}, \citet{rathinavel.bayesian_QMC_sobol}, and \citet{rathinavel.bayesian_QMC_thesis} showed that using certain LD sequences paired with matching kernels enables the prohibitive $\calO(n^3)$ cost of standard GP fitting to be reduced to $\calO(n \log n)$, including optimization of GP kernel hyperparameters. Their method draws on earlier developments by \citet{zeng.spline_lattice_digital_net,zeng.spline_lattice_error_analysis} which show that pairing LD lattices or digital nets with shift-invariant or digitally-shift-invariant kernels creates Gram matrices whose eigendecomposition can be written in terms of the fast Fourier transform (FFT) \citep{cooley1965algorithm} and fast Walsh--Hadamard transform (FWHT) \citep{fino.fwht} respectfully. Moreover, these fast Bayesian cubature constructions yield approximations (posterior integral means) which are equal to the sample mean of the function evaluations, as is the case for standard QMC rules (provided the fast GPs are parameterized with a constant prior mean which is chosen to minimize certain loss functions such as the marginal log likelihood). 

Other error estimation techniques exist for QMC with a single LD sequence. One alternative method is to track the decay of complex exponential Fourier coefficients when using lattices \citep{cubqmclattice} or Walsh coefficients using digital nets \citep{hickernell.adaptive_dn_cubature}. Discrete approximations to these coefficients may be computed at $\calO(n \log n)$ cost using the FFT and FWHT respectfully, and an error bound in terms of these discrete coefficients may be derived for a data-driven cone of compatible functions. Unified treatments of these methods are given by \citet{adaptive_qmc} and \citet{ding2018adaptive}. Applying such methods to MLQMC remains a promising topic for future work.

We propose to apply fast Bayesian cubature to MLQMC with only a single LD sequence per level. Specifically, we will fit independent GP models to the difference-of-simulations on each level and use an iterative-doubling algorithm similar to the standard greedy algorithm for MLQMC with multiple randomizations. Aside from providing more accurate estimates by avoiding the need for multiple randomizations, the Bayesian framework also enables a principled approach to sample allocation decisions. The key observation is that the Bayesian QMC error estimate (the GP posterior integral variance) is only a function of the GP hyperparameters and the LD sampling locations, \emph{not} the simulation values at the LD nodes. At a given iteration, if we assume the GP hyperparameters have been estimated exactly, then one may forecast what the error estimate will be after doubling the sample size on any level without actually having to evaluate the simulations at those LD points. We propose a scheme which chooses to double the sample size at the level with the maximum projected decrease in the aggregate MLQMC error estimate for the cost. 

The novel contributions in this paper are to: 
\begin{enumerate}
    \item Propose a Bayesian MLQMC method which, to the best of our knowledge, is the first MLQMC method which avoids the need for multiple randomizations on each level. 
    \item Develop a novel sample allocation scheme for the Bayesian MLQMC framework which exploits the ability to forecast future error estimates through GP posterior variances.
    \item Put forward a new digitally-shift-invariant (DSI) kernel with adaptive smoothness for use with fast digital net GPs. This new kernel takes a hyperparameter-weighted sum of existing higher-order smoothness DSI kernels from \citet{sorokin.2025.ld_randomizations_ho_nets_fast_kernel_mats}.
    \item Run numerical experiments validating the performance of single-level and multilevel Bayesian cubature for both digital nets and lattices.
\end{enumerate}

The remainder of this article is organized as follows. \Cref{sec:methods} details our methods. We describe the existing IID MLMC method in \Cref{sec:mlmc}, the existing MLQMC method with multiple randomizations in \Cref{sec:mlqmc}, and the proposed fast Bayesian MLQMC method in \Cref{sec:bmlqmc}. Algorithms are provided for each of these three methods. Next, \Cref{sec:numerical_experiments} gives numerical experiments comparing the three algorithms for both single-level problems (\Cref{sec:examples_single_level}) and multilevel problems with a fixed number of levels (\Cref{sec:examples_multilevel}). The two QMC algorithms will consider both digital net and lattice LD point sets. Finally, \Cref{sec:conclusion} provides concluding remarks.  

\section{Methods} \label{sec:methods}

Assume we have access to quantities of interest $Q_\ell: [0,1]^d \to \bbR$ where $\ell \in \{1,\dots,L\}$ indexes the fidelity of the simulation, with larger $\ell$ corresponding to higher-fidelity simulations. Sometimes the dimension $d$ varies with the level so $Q_\ell: [0,1]^{d_\ell} \to \bbR$; for example, in our option pricing examples in \Cref{sec:option_pricing}, $d_\ell$ will be the  number of monitoring times of an asset price with $d_\ell$ increasing in $\ell$. In such case, we take $d = \max(d_1,\dots,d_L)$ and subset dimensions accordingly. We will also assume the number of levels is fixed, although extensions to adaptive multilevel schemes, where the highest-fidelity level is determined by the remaining bias, should be straightforward \citep{giles.MLMC_path_simulation,giles2015multilevel}.

We are interested in approximating the expectation of the maximum-fidelity simulation $\nu := \bbE[Q_L(\bX)]$ subject to $\bX \sim \calU[0,1]^d$. For problems with nonuniform stochasticity, one may employ a change of variables to produce a compatible form.
As is standard for MLMC methods, we will expand the above quantity of interest as a telescoping sum 
\begin{equation} \label{eq:tele_mlmc}
    \nu = \bbE[Q_L(\bX)] = \sum_{\ell=1}^L \bbE[Q_\ell(\bX) - Q_{\ell-1}(\bX)] = \sum_{\ell=1}^L \bbE[Y_\ell(\bX)] = \sum_{\ell=1}^L \mu_\ell
\end{equation}
where we set $Y_\ell := Q_\ell-Q_{\ell-1}$, $\mu_\ell := \bbE[Y_\ell]$, and $Q_0 = 0$ so $Y_1 = Q_1$.
We will denote by $C_\ell$ the cost of evaluating $Y_\ell$ and by $V_\ell := \bbV[Y_\ell(\bX)]$ the variance of the $\ell^\mathrm{th}$ difference. 

\subsection{Multilevel Monte Carlo with Independent Points} \label{sec:mlmc}

Classic Monte Carlo methods approximate the true mean of an expectation by the $n$-sample mean at IID (independent and identically distributed) sampling locations. Standard MLMC methods \citep{giles.MLMC_path_simulation,giles2015multilevel} use IID samples on each level, 
$$\bx_{1,0},\dots,\bx_{1,n_1-1},\dots,\bx_{L,0},\dots,\bx_{L,n_L-1} \simiid \calU[0,1]^d,$$
to create the estimate 
\begin{equation} \label{eq:mlmc}
    \hnu = \sum_{\ell=1}^L \hmu_\ell \qqtqq{where} \hmu_\ell = \frac{1}{n_\ell} \sum_{i=0}^{n_\ell-1} Y_\ell(\bx_{\ell,i}).
\end{equation}
For MLMC with an adaptive number of levels, one often decomposes the mean-squared error (MSE) of \eqref{eq:mlmc} into the variance and squared bias via $\mathrm{MSE}(\hnu) = \bbE[(\hnu - \nu)^2] = \bbV[\hnu] + (\bbE[\hnu] - \nu)^2$. In our setting, $\hnu$ is unbiased for our target quantity $\nu$, the expectation of the highest-fidelity model. This implies the MSE is exactly the variance of the estimator, 
$$\mathrm{MSE}(\hnu) = \bbV[\hnu] = \sum_{\ell=1}^L \frac{V_\ell}{n_\ell}.$$

For a given error tolerance on $\bbV(\hnu)$, standard MLMC methods will optimally allocate $n_\ell \propto \sqrt{V_\ell/ C_\ell}$. As $V_\ell$ is not known, it is typically approximated by an unbiased estimator from an initial pilot sample. For ease of presentation and consistency with the QMC schemes to be presented in \Cref{sec:mlqmc} and \Cref{sec:bmlqmc}, we will use an iterative-doubling version of this IID MLMC scheme as detailed in \Cref{alg:mlmc}. Although this greedy MLMC algorithm may yield suboptimal allocations, it will approximate the cost of the optimal algorithm up to a constant factor. 

\begin{algorithm}[!ht]
    \caption{\texttt{MC}: Multilevel Monte Carlo with Independent Points}
    \label{alg:mlmc}
    \begin{algorithmic}
        \Require $N>0$ \Comment{the budget}
        \Require $C_1,\dots,C_L > 0 $ \Comment{the cost of evaluating $Y_1,\dots,Y_L$ respectively}
        \Require $n_1^\mathrm{next},\dots,n_L^\mathrm{next} \in \bbN$ satisfying $\sum_{\ell=1}^L n_\ell^\mathrm{next} C_\ell \leq N$ \Comment{the initial sample sizes}
        \State $n_\ell \gets 0$ for $\ell \in \{1,\dots,L\}$ \Comment{the number of model evaluations on each level}
        \State $\calL \gets \{1,\dots,L\}$ \Comment{the set of levels to update}
        \While{true}
            \State Generate $\bx_{\ell,i} \sim \calU[0,1]^d$ for $\ell \in \calL$ and $n_\ell \leq i < n_\ell^\mathrm{next}$ all independently
            \State Evaluate $Y_\ell(\bx_{\ell,i})$ for $\ell \in \calL$ and $n_\ell \leq i < n_\ell^\mathrm{next}$
            \State $\hmu_\ell \gets \frac{1}{n_\ell} \sum_{i=0}^{n_\ell-1} Y_\ell(\bx_{\ell,i})$ for $\ell \in \calL$ \Comment{estimate of $\mu_\ell$}
            \State $\hsigma_\ell^2 \gets \frac{1}{n_\ell-1} \sum_{i=0}^{n_\ell-1} (Y_\ell(\bx_{\ell,i}) - \hmu_\ell)^2$ for $\ell \in \calL$\Comment{estimate of $V_\ell$}
            \State $\calL_\mathrm{feasible} \gets \{\ell \in \{1,\dots,L\}: \sum_{\ell'=1}^L C_{\ell'} n_{\ell'} + C_\ell n_\ell \leq N\}$ \Comment{feasible set of levels}
            \If{$\calL_\mathrm{feasible} = \emptyset$} break \EndIf \Comment{exit while loop if it is not within budget to double on any level}
            \State $\starhat{\ell} \gets \argmax_{\ell \in \calL_\mathrm{feasible}} \frac{\sigma_\ell^2}{n_\ell C_\ell}$ \Comment{choose the level of maximum utility} 
            \State $\calL \gets \{\starhat{\ell}\}$ and $n_{\starhat{\ell}} \gets n_{\starhat{\ell}}^\mathrm{next}$ and $n_{\starhat{\ell}}^\mathrm{next} \gets 2n_{\starhat{\ell}}$ \Comment{double the sample size on the chosen level}
        \EndWhile
        \State $\hnu \gets \sum_{\ell=1}^L \hmu_\ell$ \Comment{estimate of $\nu$}
        \State $\sigma^2 \gets \sum_{\ell=1}^L \hsigma_\ell^2/n_\ell$ \Comment{estimate of $\bbV\left[\hnu\right]$}
        \\ \Return $\hnu,\sigma,\{n_\ell\}_{\ell=1}^L$ \Comment{the estimate, its standard error, and number of samples per level}
    \end{algorithmic}
\end{algorithm}

\subsection{Multilevel quasi-Monte Carlo with Replications} \label{sec:mlqmc}

\begin{figure}[!ht]
    \centering
    \includegraphics[width=1\textwidth]{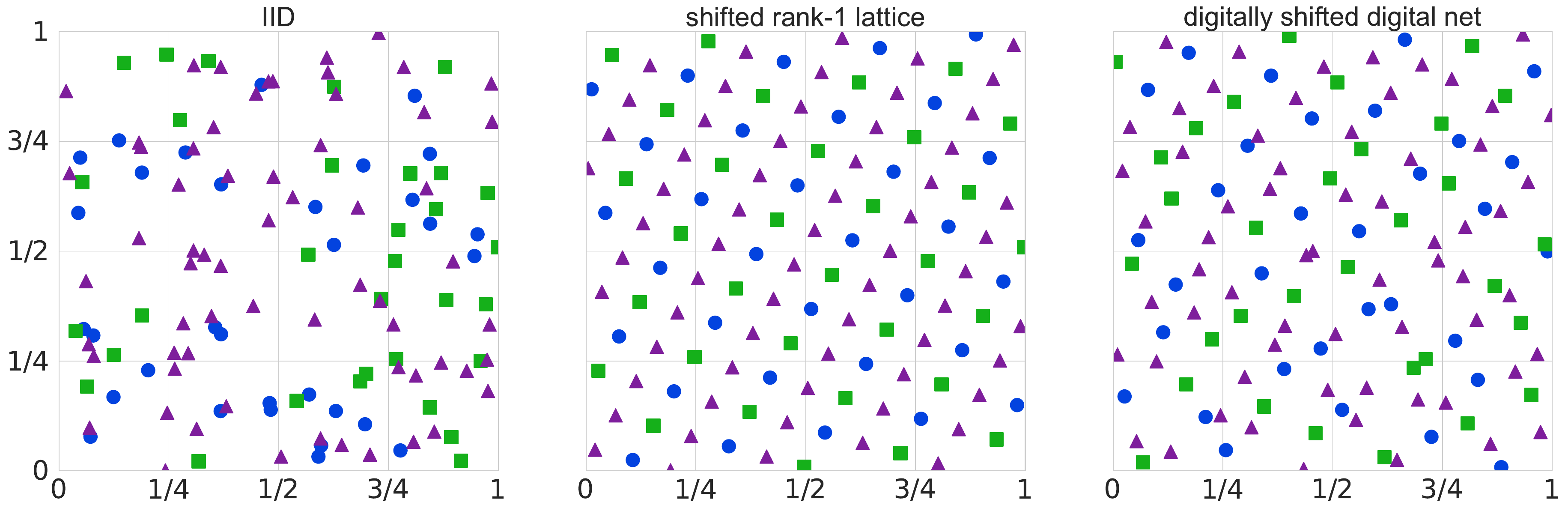}
    \caption{IID (independent identically distributed) points alongside an LD (low-discrepancy) lattice and digit net. The lattice has been randomly shifted (see \Cref{def:lattices}) and the digital net has a linear matrix scramble (LMS) and random digital shift (see \Cref{def:dnets}). IID points have gaps and clusters while the dependent LD points fill the space more evenly. Increasing the LD sample sizes through powers of two fills in the gaps left by previous points. The points are colored as follows: blue circles for the first 32 points, green squares for the next 32, and purple triangles for the subsequent 64.} 
    \label{fig:points}
\end{figure}

Quasi-Monte Carlo (QMC) methods replace the IID points from standard Monte Carlo methods with low-discrepancy (LD) sequences which more evenly cover the unit cube \citep{hickernell.qmc_what_why_how,dick.high_dim_integration_qmc_way,owen.mc_book,owen.mc_book_practical,niederreiter.qmc_book,caflisch1998monte,dick.digital_nets_sequences_book,dick2022lattice}. Certain randomizations of LD sequences retain their space filling properties while enabling QMC error estimation. In this paper we will consider the following two LD sequences with random shifts. These are visualized in \Cref{fig:points} alongside IID points. 

\begin{definition}[Shifted rank-1 lattice] \label{def:lattices}
    For a fixed generating vector $\bg \in \bbN^d$, the rank-1 lattice $(\bz_i)_{i \geq 0} \subset [0,1)^d$ in the extensible radical inverse order is 
    $$\bz_i = v(i) \bg \mod 1, \qquad v(0)=0, \qquad \bv(i) = \sum_{p=0}^{\lfloor \log_2(i) \rfloor} \mi_p 2^{-p-1}, \qquad i=\sum_{p=0}^{\lfloor \log_2(i) \rfloor } \mi_p 2^p,$$
    where $v(i)$ is the van der Corput sequence~\citep{vandercorput}. For a shift $\bDelta \in [0,1)^d$, the shifted rank-1 lattice $(\bx_i)_{i \geq 0}$ is 
    $$\bx_i = \bz_i \oplus \bDelta \qqtqq{where} \ba \oplus \bb = (\ba + \bb) \mod 1 \quad \forall\;\ba,\bb \in [0,1]^d$$
    with modulo taken element-wise.
\end{definition}

\begin{definition}[Digitally-shifted digital net in base $2$] \label{def:dnets}
    For generating matrices $\mG \in [0,1)^{d \times \infty}$ with $0^\mathrm{th}$ column $\bg_0$ and $p^\text{th}$ column $\bg_p$, the digital net $(\bz_i)_{i \geq 0} \subset [0,1)^d$ in the extensible radical inverse order is  
    $$\bz_i = \bigoplus_{p=0}^{\lfloor\log_2(i)\rfloor} \mi_p \bg_p, \qquad \ba \oplus \bb := (a_1 \oplus b_1,\dots,a_d \oplus b_d) \quad \forall\; \ba,\bb \in [0,1)^d,$$
    where, for $a,b \in [0,1)$, we define $\oplus$ to take the exclusive or (XOR) of binary expansions, i.e.,
    $$a \oplus b = \sum_{p \geq 0} ((\ma_p + \mb_p) \mod 2) 2^{-p}, \qquad a = \sum_{p \geq 0} \ma_p 2^{-p}, \qquad b = \sum_{p \geq 0} \mb_p 2^{-p}.$$
    For a digital-shift $\bDelta \in [0,1)^d$, the digitally-shifted digital net $(\bx_i)_{i \geq 0}$ is
    $$\bx_i = \bz_i \oplus \bDelta.$$
\end{definition} 

The definition of the shift operator $\oplus$ is different for lattices and digital nets. Lattices and digital nets have been well studied, see the books by \citet{dick2022lattice} and \citet{dick.digital_nets_sequences_book} respectfully. While we present and use base $2$ digital nets, which permit efficient implementation with the XOR (exclusive or) operation, analogous definitions of digital nets in arbitrary prime base are also available. We call $\mG \in [0,1)^{d \times \infty}$ ``generating matrices'' as one may think of each row of $\mG$ as a $\infty \times \infty$ matrix of zeros and ones of binary expansions of each element. We technically require that the binary expansion of each element of $\mG$ does \emph{not} end in an infinite tail of ones, but this is not an issue in our finite precision implementation. The presented construction allows for linear matrix scrambling (LMS) of the generating matrices $\mG$ \citep{MATOUSEK1998527}. Nested uniform scrambling (Owen scrambling) \citep{owen.variance_alternative_scrambles_digital_net} is also supported for the QMC methods in this section which use independent randomizations, but it is \emph{not} compatible with the Bayesian methods we will explore in \Cref{sec:bmlqmc}.  

When $\bDelta \sim \calU[0,1]^d$, we have $\bx_i \sim \calU[0,1]^d$ for both the randomly shifted rank-1 lattices in \Cref{def:lattices} and the randomly digitally-shifted digital nets in \Cref{def:dnets}. The multilevel QMC (MLQMC) method from \citep{giles.mlqmc_path_simulation} uses $R$ independent shifts on each level
$$\bDelta_{1,1},\dots,\bDelta_{1,R},\dots,\bDelta_{L,1},\dots,\bDelta_{L,r} \simiid \calU[0,1)^d$$
to construct $LR$ independent randomizations (replications) of an LD point sets 
$$(\bx_{\ell,r,i})_{i=0}^{n_\ell-1}, \qquad \bx_{\ell,r,i} := \bz_i \oplus \bDelta_{\ell,r}, \qquad 1 \leq \ell \leq L, \quad 1 \leq r \leq R.$$
One then replaces $\hnu$ in \eqref{eq:mlmc} by an approximation derived from the $LR$ independent sample means $\tmu_{\ell,r}$:
\begin{equation} \label{eq:mlqmc} 
    \hnu = \sum_{\ell=1}^L \hmu_\ell \qqtqq{where} \hmu_\ell = \frac{1}{R} \sum_{r=1}^R \tmu_{\ell,r} \qqtqq{and} \tmu_{\ell,r} = \frac{1}{n_\ell} \sum_{i=0}^{n_\ell-1} Y_\ell(\bx_{\ell,r,i}).
\end{equation}
Then the variance of $\hnu$ with respect to a uniform random shift $\bDelta$ is  
$$\bbV\left[\hnu\right] = \sum_{\ell=1}^L \frac{\tV_{\ell,n_\ell}}{R} \qqtqq{where} \tV_{\ell,n_\ell} = \bbV\left[\frac{1}{n_\ell}\sum_{i=0}^{n_\ell-1} Y_\ell(\bz_i \oplus \bDelta)\right] \qqtqq{with} \bDelta \sim \calU[0,1]^d.$$
We expect each $\tV_{\ell,n_\ell}$ to approach $0$ as $n_\ell \to \infty$. Rather than increasing the number of randomizations $R$, it will be more efficient to increase the number of LD point $n_\ell$ and exploit the space filling designs of extensible LD sequences. This constitutes the main drawback of randomizations: computational effort that could otherwise be spent on improving the accuracy of the estimate $\hnu$ is wasted on computing a statistical error estimate of $\tV_{\ell,n_\ell}$. 

We present the standard iterative-doubling multilevel QMC method in \Cref{alg:mlqmc}, which exploits the fact that the extensible lattices and base $2$ digital sequences we consider achieve desirable space filling properties at samples sizes which are powers of $2$. In the trivial case when $n_\ell = 1$, the estimates $\hmu_\ell$ in \eqref{eq:mlmc} and \eqref{eq:mlqmc} have the same distribution. Of course one could use a different number of randomizations $R_\ell$ on each level, but this is not common in the literature and is not considered here. We also choose to fix the generating vector $\bg$ for lattices in \Cref{def:lattices} and fix the generating matrices $\mG$ for digital nets in \Cref{def:dnets} across both levels and randomizations.

\begin{algorithm}[!ht]
    \caption{\texttt{RQMC}: Multilevel Quasi-Monte Carlo With Replications}
    \label{alg:mlqmc}
    \begin{algorithmic}
        \Require $N>0$ \Comment{the budget}
        \Require $C_1,\dots,C_L > 0 $ \Comment{the cost of evaluating $Y_1,\dots,Y_L$ respectively}
        \Require $R>0$ \Comment{the fixed number of randomizations on each level, we use $R=8$.} 
        \Require $n_1^\mathrm{next},\dots,n_L^\mathrm{next} \in \bbN$ powers of two satisfying $R \sum_{\ell=1}^L n_\ell^\mathrm{next} C_\ell \leq N$ \Comment{the initial sample sizes}
        \Require A generating vector $\bg \in \bbN^d$ to use lattices from \Cref{def:lattices} or \\ generating matrices $\mG \in [0,1)^{d \times \infty}$ to use base $2$ digital nets from \Cref{def:dnets}. 
        \State Generate $\bDelta_{1,1},\dots,\bDelta_{1,R},\dots,\bDelta_{L,1},\dots,\bDelta_{L,r} \simiid \calU[0,1)^d$ \Comment{random shifts}
        \State $n_\ell \gets 0$ for $\ell \in \{1,\dots,L\}$ \Comment{the number of evaluations per randomization}
        \State $\calL \gets \{1,\dots,L\}$ \Comment{the set of levels to update}
        \While{true}
            \State Generate $x_{\ell,r,i} = \bz_i \oplus \bDelta_{\ell,r}$ for $\ell \in \calL$ and $1 \leq r \leq R$ and $n_\ell \leq i < n_\ell^\mathrm{next}$ \Comment{\Cref{def:lattices,def:dnets}}
            \State Evaluate $Y_\ell(\bx_{\ell,r,i})$ for $\ell \in \calL$ and $1 \leq r \leq R$ and $n_\ell \leq i < n_\ell^\mathrm{next}$
            \State $\tmu_{\ell,r} \gets \frac{1}{n_\ell} \sum_{i=0}^{n_\ell-1} Y_\ell(\bx_{\ell,r,i})$ for $\ell \in \calL$ and $1 \leq r \leq R$ \Comment{per-randomization estimate of $\mu_\ell$}
            \State $\hmu_\ell \gets \frac{1}{R} \sum_{r=1}^{R} \tmu_{\ell,r}$ for $\ell \in \calL$ \Comment{aggregate estimate of $\mu_\ell$}
            \State $\tsigma_\ell^2 \gets \frac{1}{R-1} \sum_{r=1}^{R} (\tmu_{\ell,r} - \hmu_\ell)^2$ for $\ell \in \calL$ \Comment{estimate of $\tV_{\ell,n_\ell}$}
            \State $\calL_\mathrm{feasible} \gets \{\ell \in \{1,\dots,L\}: \sum_{\ell'=1}^L R C_{\ell'} n_{\ell'} + R C_\ell n_\ell \leq N\}$ \Comment{feasible set of levels}
            \If{$\calL_\mathrm{feasible} = \emptyset$} break \EndIf \Comment{exit while loop if it is not within budget to double on any level}
            \State $\starhat{\ell} \gets \argmax_{\ell \in \calL_\mathrm{feasible}} \frac{\tsigma_\ell^2}{R n_\ell C_\ell}$ \Comment{choose the level of maximum utility}
            \State $\calL \gets \{\starhat{\ell}\}$ and $n_{\starhat{\ell}} \gets n_{\starhat{\ell}}^\mathrm{next}$ and $n_{\starhat{\ell}}^\mathrm{next} \gets 2n_{\starhat{\ell}}$ \Comment{double the sample size on the chosen level}
        \EndWhile
        \State $\hnu \gets \sum_{\ell=1}^L \hmu_\ell$ \Comment{estimate for $\nu$}
        \State $\tsigma^2 \gets \sum_{\ell=1}^L \tsigma_\ell^2/R$ \Comment{estimate of $\bbV[\hnu]$}    
        \\ \Return $\hnu,\tsigma,\{R n_\ell\}_{\ell=1}^L$ \Comment{the estimate, its standard error, and number of samples per level}
    \end{algorithmic}
\end{algorithm}

\subsection{Fast Bayesian Multilevel Quasi-Monte Carlo Without Replications} \label{sec:bmlqmc}

We will assume $Y_\ell$ is a Gaussian process (GP) \citep{rasmussen.gp4ml} and, while not necessarily justified theoretically, we make the modeling assumption that $Y_1,\dots,Y_L$ are independent GPs. Suppose $Y$ is a GP with prior mean $M:[0,1]^d \to \bbR$ and symmetric positive definite (SPD) covariance kernel $K:[0,1]^d \times [0,1]^d \to \bbR$. Here we have dropped the subscript $\ell$ for ease of presentation. Suppose we have sampling locations $\mX = (\bx_i)_{i=0}^{n-1} \subset [0,1)^d$ and function evaluations $\bY := (Y(\bx_i))_{i=0}^{n-1} \in \bbR^{n}$. In the multilevel setting, the prior mean, kernel, and sampling locations may be different on each level. We will also use $\bM := (M(\bx_i))_{i=0}^{n-1} \in \bbR^n$ to denote the vector of mean evaluations, $\bK(\bx) = (K(\bx,\bx_i))_{i=0}^{n-1}$ to denote the vector of kernel evaluations, and $\mK = (K(\bx_i,\bx_{i'}))_{i,i'=0}^{n-1}$ to denote the Gram matrix of pairwise kernel evaluations. Then the posterior (conditional) distribution is a GP with mean and covariance 
\begin{align*}
    \bbE[Y(\bx) | \mX,\bY] &= M(\bx) + \bK^\intercal(\bx) \mK^{-1} (\bY-\bM) \qquad\text{and} \\
    \Cov[Y(\bx),Y(\bx') | \mX] &= K(\bx,\bx') - \bK^\intercal(\bx) \mK^{-1} \bK(\bx')
\end{align*}
for $\bx,\bx' \in [0,1]^d$. Notice the posterior covariance depends on the sampling locations $\mX$ but \emph{not} on the function evaluations $\bY$. 

Often the kernel $K$ and prior mean $M$ will depend on hyperparameters $\btheta$ which are chosen to optimize some loss function such as the negative marginal log likelihood (NMLL) loss, the (generalized) cross validation (G)CV loss, or a full Bayesian loss. In this paper we will choose $\btheta$ to minimize the NMLL loss which, up to a constant independent of $\btheta$, may be written as  
\begin{equation}
    L(\btheta) =  \log \lvert \mK \rvert + (\bY - \bM) \mK^{-1} (\bY - \bM)
    \label{eq:nmll}
\end{equation}
with an implicit dependence of $\bM$ and $\bK$ on $\btheta$. For example, the squared exponential kernel $K(\bx,\bx') = \gamma \exp(- \sum_{j=1}^d (x_j-x'_j)^2/(2 \eta_j^2))$ has a global scaling factor $\gamma>0$ and lengthscales $\bEta = (\eta_1,\dots,\eta_d)>0$, and 
we will consider a constant prior mean $M(\bx) = \tau$ for all $\bx \in [0,1]^d$, so $\btheta = \{\gamma,\bEta,\tau\}$.

Assuming evaluation of the kernel $K$ and prior mean $M$ cost $\calO(d)$, fitting a GP will generally require $\calO(n^3+n^2d)$ computations and $\calO(n^2)$ storage and as one needs to evaluate, store, invert, and compute the determinant of the $n \times n$ SPD Gram matrix $\mK$. After evaluating and storing $\mK$, standard practice is to compute its Cholesky decomposition at $\calO(n^3)$ cost and then perform back-substitutions to solve the linear system $\mK^{-1}(\bY - \bM)$ at $\calO(n^2)$ cost. If $\mK$ is well conditioned, then one may use a preconditioned conjugate gradient (PCG) method to solve the linear system at only $\calO(n^2)$ cost \citep{gardner.gpytorch_GPU_conjugate_gradient}. PCG also scales nicely on GPUs which expose efficient matrix multiplication routines. Even so, PCG and other GP acceleration routines often require at least $\calO(d n^2)$ computations and $\calO(n^2)$ storage. These requirements are prohibitive to our setting, where often tens of thousands of samples are collected on the coarser levels. We will use different GP acceleration methods that force specific structure into $\mK$ which enables reduced computation and storage requirements. These methods are similar to using regular grids with product kernels to create Kronecker or Toeplitz structure \citep{wilson2014covariance,saatcci2012scalable,gardner2018product,wilson2015kernel}, but allow us to use LD point sets for high-dimensional QMC approximations. 

\subsubsection{Fast Gaussian Processes using Low-Discrepancy Sequences and Matching Kernels} 

We will use set of GP acceleration techniques which only require $\calO(n \log n+nd)$ computations and $\calO(nd)$ storage. These techniques use LD QMC point sets paired with special kernel forms as originally proposed by \citet{zeng.spline_lattice_digital_net, zeng.spline_lattice_error_analysis} and recently applied to both fast automatic Bayesian cubature \citep{rathinavel.bayesian_QMC_lattice,rathinavel.bayesian_QMC_sobol,rathinavel.bayesian_QMC_thesis} and fast GP regression \citep{kaarnioja.kernel_interpolants_lattice_rkhs,kaarnioja.kernel_interpolants_lattice_rkhs_serendipitous,sorokin.gp4darcy,sorokin.fastgps_probnum25}. Such pairings yield Gram matrices $\mK$ whose structure may be exploited for fast kernel computations. Specifically, 
\begin{enumerate}
    \item pairing a shifted rank-1 integration lattice in radical inverse order from \Cref{def:lattices} with a shift-invariant (SI) kernel to be defined in \Cref{def:si_kernel} will produce $\mK$ which is diagonalizable by the discrete Fourier transform applied in bit-reversed order, and 
    \item pairing a digitally-shifted base $2$ digital net in radical inverse order from \Cref{def:dnets} with a digitally-shift-invariant (DSI) kernel to be defined in \Cref{def:dsi_kernel} will produce $\mK$ which is diagonalizable by the discrete Walsh--Hadamard transform. 
\end{enumerate}
Applying the discrete Fourier transform in bit-reversed order or the discrete Walsh--Hadamard transform to $n$ points requires only $\calO(n \log n)$ computations by respectively using the fast Fourier transform \citep{cooley1965algorithm} in bit-reversed order (FFTBR) and the fast Walsh--Hadamard Transform (FWHT) \citep{fino.fwht} algorithms.

The key observation underlying these fast GP methods is that, under either of the special pairings enumerated above, the Gram matrix $\mK$ has an eigendecomposition
$$\mK = \mE \mLambda \overline{\mE}$$
where, for any $\ba \in \bbC^n$,  $\tba := \overline{\mE} \ba$ is the FFTBR or FWHT of $\ba$ which can be computed at $\calO(n \log n)$ cost (with $\overline{\mE}$ denoting the complex conjugate of $\mE$ applied elementwise). This enables us to compute eigenvalues at $\calO(n \log n)$ cost using 
$$\blambda = \mLambda \bone = \sqrt{n} \; \overline{\mE} \mE \mLambda \overline{\mE}_{:,0} = \sqrt{n} \; \overline{\mE} \mK_{:,0}$$
where we assume $\mE$ is scaled to be orthonormal, i.e., $\mE \overline{\mE} = \mI$, and $\overline{\mE}_{:,0} = \bone /\sqrt{n}$ (the $0^\mathrm{th}$ column of $\overline{\mE}$). Therefore, it is sufficient to only evaluate $\mK_{:,0}$ (the $0^\mathrm{th}$ column of $\mK$) and only store the eigenvalues $\blambda$, so our storage costs are reduced to $\calO(n)$. One may also solve a linear system in $\mK$ for any $\ba$ at $\calO(n \log n)$ cost using 
$$\mK^{-1} \ba = \mE (\tba ./ \blambda)$$
where $./$ denote elementwise division. The determinant $\lvert \mK \rvert = \lvert \prod_{i=0}^{n-1} \lambda_i \rvert$ can also be computed at $\calO(n \log n)$ cost. 

The multivariate product kernel we present in \Cref{def:prod_kernel} will combine univariate SI and DSI kernels from \Cref{def:si_kernel,def:dsi_kernel} into multivariate SI and DSI kernels respectively. 

\begin{definition}[Product kernels] \label{def:prod_kernel}
    Our SPD SI/DSI kernels $K:[0,1)^d \times [0,1)^d \to \bbR$ take the product form 
    \begin{equation}
        K(\bx,\bx') = \gamma \calK(\bx,\bx') \qqtqq{with} \calK(\bx,\bx') = \prod_{j=1}^d (1+\eta_j R(x_j,x_j')) \qquad \forall\; \bx,\bx' \in [0,1]^d.
        \label{eq:product_kernel}
    \end{equation}
    Here $R:[0,1) \to \bbR$ is a univariate SI/DSI kernel, $\gamma > 0$ is a global scaling parameter, and $\bEta > 0$ are marginal scaling factors which we will also call lengthscales. 
\end{definition}

\begin{definition}[Shift-invariant (SI) product kernels] \label{def:si_kernel}
    Our SI kernels take the form of \eqref{eq:product_kernel} using univariate kernels
    \begin{equation}
        \begin{aligned} 
        R(x,x') &= \tR_\alpha(x \oplus x') \qtq{with} \tR_\alpha(x) &=  \frac{(-1)^{\alpha+1}(2 \pi)^{2 \alpha}}{(2\alpha)!} B_{2\alpha-\beta-\beta'}(x), \qquad \alpha \in \bbN.
        \end{aligned}
        \label{eq:si_kernels}
    \end{equation}
    Here $B_p$ denotes the $p^\mathrm{th}$ Bernoulli polynomial and $x \oplus x' = (x+x') \mod 1$ as used in \Cref{def:lattices} of lattices. 
\end{definition}

These SI kernels are periodic and correspond to weighted Korobov spaces \citep{kaarnioja.kernel_interpolants_lattice_rkhs,kaarnioja.kernel_interpolants_lattice_rkhs_serendipitous,cools2021fast,cools2020lattice,sloan2001tractability,kuo2004lattice} with larger $\alpha \in \bbN$ corresponding to greater smoothness. In our numerical experiments, we will use the $\alpha=1$ kernel to assume as little periodicity as possible.

\begin{definition}[Digitally-shift-invariant (DSI) product kernels] \label{def:dsi_kernel}
    Our DSI kernels take the form of \eqref{eq:product_kernel} using a weighted sum of univariate kernels with hyperparameter weights $\bbeta >0$ so that 
    \begin{equation}
        \begin{aligned} 
        R(x,x') &:= \tR(x \oplus x') = \beta_1 \tR_1(x,x') + \beta_2 \tR_2(x,x') + \beta_3 \tR_3(x,x') + \beta_4 \tR_4(x,x') \qquad\text{with } \\
        \tR_\alpha(x) &= \begin{cases} 6\left(1-\frac{1}{2} t_1(x)\right), & \alpha=1 \\ -1+-\beta(x) x + \frac{5}{2}\left[1-t_1(x)\right], & \alpha = 2 \\ -1+\beta(x)x^2-5\left[1-t_1(x)\right]x+\frac{43}{18}\left[1-t_2(x)\right], & \alpha = 3 \\ -1 -\frac{2}{3}\beta(x)x^3+5\left[1-t_1(x)\right]x^2 - \frac{43}{9}\left[1-t_2(x)\right]x \\
        \quad +\frac{701}{294}\left[1-t_3(x)\right]+\beta(x)\left[\frac{1}{48} \sum_{a=1}^\infty \frac{\mx_a}{2^{3(a-1)}} - \frac{1}{42}\right], & \alpha = 4 \end{cases}.
        \end{aligned}
        \label{eq:dsi_kernels}
    \end{equation}
    Here $\beta(x) = - \lfloor \log_2(x) \rfloor$, $t_\nu(x) = 2^{-\nu \beta(x)}$, and $x \oplus x' = \sum_{a=1}^\infty ((\mx_a + \mx_a') \mod 2) 2^{-(a+1)}$ is the exclusive or (XOR) between binary expansions of $x = \sum_{a =0}^\infty \mx_a 2^{-a}$ and $x' = \sum_{a=0}^\infty \mx_a' 2^{-a}$ as used in \Cref{def:dnets} of digital nets. 
\end{definition}

For these DSI kernels, the $\alpha=1$ form is due to \citet{dick.multivariate_integraion_sobolev_spaces_digital_nets}, and the higher-order $\alpha \geq 2$ kernel forms are due to \citet{sorokin.2025.ld_randomizations_ho_nets_fast_kernel_mats} with some recent numerical experiments given by \citet{sorokin.fastgps_probnum25}. Despite having discontinuous kernels, the reproducing kernel Hilbert spaces (RKHSs) corresponding to the $\alpha \geq 2$ forms contain Sobolev spaces of smooth, non-periodic functions. We are the first to consider weighted sums of these univariate DSI kernels, which enables optimizing the weights $\bbeta$ to select the kernel's smoothness. 

Assuming a constant prior mean $M(\bx) = \tau$ for all $\bx \in [0,1]^d$, we may rewrite the NMLL loss in \eqref{eq:nmll} as
$$L(\btheta) = \sum_{i=0}^{n-1} \log \lvert \lambda_i \rvert + (\bY - \tau \bone)^\intercal \mE \mLambda^{-1} \overline{\mE} (\bY - \tau \bone)$$
where $\bone$ is the length $n$ vector of ones. The SI kernels in \Cref{def:si_kernel} have hyperparameters $\btheta = \{\gamma,\bEta,\tau\}$, and the DSI kernels in \Cref{def:dsi_kernel} have hyperparameters $\btheta = \{\gamma,\bEta,\tau,\bbeta\}$. The NMLL-minimizing prior mean $\tau$ is exactly the sample mean $\tau = 1/n \sum_{i=0}^{n-1} Y(\bx_i)$ considered throughout the QMC literature. An exact form for the NMLL-minimizing global scaling factor $\gamma$ is also available \citep{rathinavel.bayesian_QMC_thesis}.

\subsubsection{Fast Bayesian Cubature} 

Bayesian cubature \citep{briol2019probabilistic,o1991bayes,rasmussen2003bayesian,briol.frank_wolfe_bayesian_quadrature} uses the fact that, since integration is a linear functional, the posterior distribution of the mean $\mu = \bbE_\bX[Y(\bX)]$ with $\bX \sim \calU[0,1]^d$ is a univariate Gaussian with mean and variance 
\begin{equation} \label{eq:post_cubature_mean_var}
    \begin{aligned}
        \hmu := \bbE[\mu | \mX,\bY] &= \int M(\bx) \D \bx + \left(\int \bK(\bx) \D \bx\right)^\intercal \mK^{-1}(\bY - \bM) \qquad\text{and} \\
        \hV_n := \bbV[\mu | \mX] &= \int \int K(\bx,\bx') \D \bx \D \bx' - \left(\int \bK(\bx) \D \bx\right)^\intercal \mK^{-1}\left(\int \bK(\bx) \D \bx\right)
    \end{aligned}
\end{equation}
where the integrals are understood to be over $[0,1]^d$ and act elementwise on vector functions. The SI kernels in \Cref{def:si_kernel} and the DSI kernels in \Cref{def:dsi_kernel} both satisfy $\int K(\bx,\bx') \D \bx' = \gamma$ for any $\bx \in [0,1]^d$. Using an SI/DSI kernel and the NMLL-minimizing constant prior mean $\tau= 1/n \sum_{i=0}^{n-1} Y(\bx_i)$, the posterior cubature mean and variance in \eqref{eq:post_cubature_mean_var} become 
\begin{equation} \label{eq:fgp_post_cubature_mean_var}
    \hmu = \frac{1}{n} \sum_{i=0}^{n-1} Y(\bx_i) \qqtqq{and} \hV_n = \gamma \left[1-\left(\frac{1}{n} \sum_{i=0}^{n-1} \calK(\bx_i,\bx_0)\right)^{-1}\right],
\end{equation}
where $\calK(\bx,\bx') = K(\bx,\bx') / \gamma = \prod_{j=1}^d \left(1+\eta_j R(x \oplus x_j')\right)$ is the unscaled kernel as in \Cref{def:prod_kernel}. 

\begin{figure}[!ht]
    \centering 
    \begin{tikzpicture}
        \draw[-,thick] (0,0) -- (4,0);
        \draw[-,thick] (0,0) -- (0,4);
        \draw[-,thick] (0.5,3.5) -- (3.5,0.5);
        \fill (0.5,3.5) circle [radius=.1];
        \fill (3.5,0.5) circle [radius=.1];
        \draw (2,2) circle [radius=.1];
        \draw[|-|,thick] (.5,0) -- (2,0);
        \draw[|-|,thick] (2,0) -- (3.5,0);
        \draw[|-|,thick] (0,.5) -- (0,2);
        \draw[|-|,thick] (0,2) -- (0,3.5);
        \draw[-,thin,dotted] (2,0) -- (2,2);
        \draw[-,thin,dotted] (0,2) -- (2,2);
        \draw[-,thin,dotted] (0.5,0) -- (0.5,3.5);
        \draw[-,thin,dotted] (0,0.5) -- (3.5,0.5);
        \draw[-,thin,dotted] (0,0.5) -- (3.5,0.5);
        \draw[-,thin,dotted] (0,3.5) -- (0.5,3.5);
        \draw[-,thin,dotted] (3.5,0) -- (3.5,0.5);
        \node at (0.5,-0.5) {$p$};
        \node at (2,-0.5) {$\log_2(\hn)$};
        \node at (3.5,-0.5) {$p+1$};
        \node at (-1.25,0.5) {$\log_2\left(\hV_{2^p}\right)$};
        \node at (-1.25,2) {$\log_2\left(\hV_\hn\right)$};
        \node at (-1.25,3.5) {$\log_2\left(\hV_{2^{p+1}}\right)$};
    \end{tikzpicture}
    \caption{Log-log linear interpolation for $\hV_\hn$ when $\hn$ is not a power of $2$ using $p = \lfloor \log_2(\hn) \rfloor$.}
    \label{fig:logloglininterp}
\end{figure}
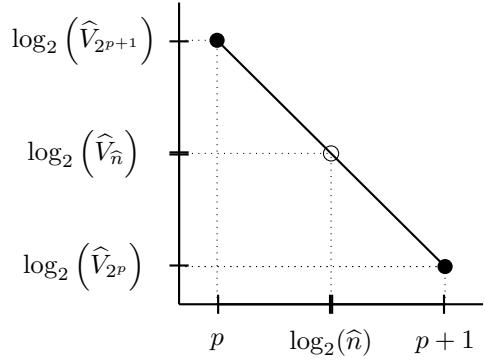

A key observation is that $\hV_n$ only depends on the points $\mX$, not the function evaluations $\bY$. This implies that, for fixed hyperparameters $\btheta$, we may compute the projected variance $\hV_\hn$ for any $\hn \geq n$ points from the given LD sequence. We prefer sample sizes $\hn$ which are powers of two as there the LD sequences we consider attain desirable uniformity properties and enable fast GP computations. While one may compute $\hV_\hn$ exactly for any $\hn$, for later utility we will define $\hV_\hn$ when $\hn$ is not a power of $2$ to be the log-log linear interpolation between surrounding powers of two so that
\begin{equation} \label{eq:logloglininterp}
    \hV_\hn := \left(\frac{\hV_{2^p}^{p+1}}{\hV_{2^{p+1}}^p}\right) \hn^{\log_2\left(\hV_{2^{p+1}}/\hV_{2^p}\right)} \qqtqq{when} 2^p < \hn < 2^{p+1} \qqtqq{for} p \in \bbN_0
\end{equation}
with $p = \lfloor \log_2(\hn) \rfloor$, see \Cref{fig:logloglininterp}.  

\subsubsection{Fast Multilevel Bayesian Cubature}

Let us now consider the multilevel Bayesian cubature setup where we fit independent GPs to each difference $Y_\ell$. We will select shifts $\bDelta_1,\dots,\bDelta_L \in [0,1)^d$ and use LD points $\bx_{\ell,i} = \bz_i \oplus \bDelta_{\ell}$ on each level. The shifts $\bDelta_1,\dots,\bDelta_L \in [0,1)^d$ are arbitrary and need not be independent as was the case for MLQMC with replication in \Cref{sec:mlqmc}. This is because the uncertainty in Bayesian cubature is quantified through the posterior variance of the Gaussian process. One may even choose the same shift on each level, $\bDelta_1 = \dots = \bDelta_L$, to use the same collocation points on each level as is common in recycling MLMC methods \citep{robbe2019recycling, kumar2017multigrid}. Different hyperparameters $\btheta_\ell$ will be fit on each level. Using \eqref{eq:fgp_post_cubature_mean_var} with level-dependent notations, we estimate $\nu$ in \eqref{eq:tele_mlmc} by the posterior cubature mean 
\begin{equation*} \label{eq:nu_bmlqmc_pcmean}
    \hnu := \bbE[\nu | (\mX_\ell,\bY_\ell)_{\ell=1}^L] = \sum_{\ell=1}^\ell \hmu_\ell \qqtqq{where} \hmu_\ell := \bbE[\mu_\ell | \mX_\ell,\bY_\ell] = \frac{1}{n_\ell} \sum_{i=0}^{n_\ell-1} Y_\ell(\bx_{\ell,i}). 
\end{equation*}
Additionally, under the assumption of independent GPs $Y_1,\dots,Y_L$, the posterior cubature variance becomes 
\begin{equation} \label{eq:nu_bmlqmc_pcvar}
    \hV_{n_1,\dots,n_L} := \bbV[\nu | (\mX_\ell)_{\ell=1}^L] = \sum_{\ell=1}^L \hV_{\ell,n_\ell} \qtq{where} \hV_{\ell,n_\ell} := \bbV[\mu_\ell | \mX_\ell] = \gamma_\ell \left[1-\left(\frac{1}{n_\ell} \sum_{i=0}^{n_\ell-1} \calK_\ell(\bx_{\ell,i},\bx_{\ell,0})\right)^{-1}\right]
\end{equation}
with $\calK_\ell$ depending on hyperparameters $\btheta_\ell \setminus \{\gamma\}$, i.e., all hyperparameters except the global scaling factor $\gamma$. If one is \emph{not} willing to assume independent GPs $Y_1,\dots,Y_L$, then the Cauchy--Schwarz inequality may be used to bound $\bbV[\nu | (\mX_\ell)_{\ell=1}^L] \leq \left(\sum_{\ell=1}^L \sqrt{\hV_{\ell,n_\ell}}\right)^2$. In our numerical experiments, we use the independent GP assumption and take the squared standard error to be $\hV_{n_1,\dots,n_L}$ in \eqref{eq:nu_bmlqmc_pcvar} as we found it provided robust error estimation across our suite of test problems.  

It is possible to develop a non-greedy sample allocation scheme using the projected variance similar to what is done for MLMC with IID points \citep{giles.mlqmc_path_simulation,giles2015multilevel}. For example, one may take an initial pilot sample, fit a GP on each level, and then uses the projected variances to determine an optimal allocation for the entire budget. However, this would rely on the assumption that the kernel hyperparameters are exactly estimated after the pilot sample, which we found to be an unreasonable assumption in our experiments. 

Instead of a complete upfront allocation, we will follow the replicated MLQMC scheme in \Cref{alg:mlqmc} and iteratively double the sample size at the level of maximum utility. The replicated QMC scheme iteratively doubles at level $\starhat{\ell} \gets \argmax_{\ell \in \calL_\mathrm{feasible}} \tsigma_\ell^2/(R n_\ell C_\ell)$ where $\tsigma_\ell^2$ is the variance of the replicated estimator. A naive utility function for Bayesian MLQMC may replace $\tsigma_\ell^2$ with the posterior cubature variance $\hV_{\ell,n_\ell}$ in \eqref{eq:nu_bmlqmc_pcvar} and use $\starhat{\ell} \gets \argmax_{\ell \in \calL_\mathrm{feasible}} \hV_{\ell,n_\ell}/(n_\ell C_\ell)$, or one may even exploit the projected variance to set $\starhat{\ell} \gets \argmax_{\ell \in \calL_\mathrm{feasible}} \hV_{\ell,2n_\ell}/(n_\ell C_\ell)$. However, we do not know the convergence rates of the GP posterior variance, so such utility functions may not be optimal. Instead, we will use the projected variance to decide which level provides the maximum error reduction for the cost. 

Our novel level selection scheme is detailed in \Cref{alg:level_select_bmlqmc}, and our complete fast Bayesian MLQMC method is given in \Cref{alg:bmlqmc}. At each iteration, our selected level will depend on the projected variance $\hV_{\ell,\hn_\ell}$ for certain $\hn_\ell$. Specifically, we compare subsequent levels in non-increasing cost-of-doubling order and iteratively select the level with the smallest projected variance for the same cost. 
Concretely, when comparing levels $\ell$ and $\ell'$, if $n_\ell C_\ell = n_{\ell'} C_{\ell'}$, i.e., the cost of doubling on each level is equivalent, 
then we choose to move forward with the level which gives the larger decrease in variance between $V_{\ell,n_\ell} - V_{\ell,2n_\ell}$ and $V_{\ell',n_{\ell'}}-V_{\ell',2n_{\ell'}}$. If $n_\ell C_\ell > n_{\ell'} C_{\ell'}$, i.e., the cost of doubling on level $\ell$ is greater than the cost of doubling on level $\ell'$, then we choose to move forward with level which gives the larger decrease in variance between $V_{\ell,n_\ell} - V_{\ell,2n_\ell}$ and $V_{\ell',n_{\ell'}} - V_{\ell',\hn_{\ell'}}$ where $\hn_{\ell'}$ is chosen to satisfy $n_\ell C_\ell = C_{\ell'}(\hn_{\ell'} - n_{\ell'})$ so doubling the sample size on level $\ell$ would require the same cost as increasing the sample size to $\hn_{\ell'}$ on level $\ell'$. As $\hn_{\ell'}$ will usually not be a power of two, we use the log-log interpolation between surrounding powers of two as in \eqref{eq:logloglininterp}. The fact that the log-log linear interpolation recovers the case when $n_\ell C_\ell = n_{\ell'} C_{\ell'}$ is reflected in the simplified presentation of \Cref{alg:level_select_bmlqmc}.

\begin{algorithm}[!ht]
    \caption{\texttt{level\_select\_BQMC}: Level selection for fast Bayesian Multilevel quasi-Monte Carlo}
    \label{alg:level_select_bmlqmc}
    \begin{algorithmic}
        \Require $\calL_\mathrm{feasible} \subseteq \{1,\dots,L\}$ with $\tL := \lvert \calL_\mathrm{feasible} \rvert>0$ elements \Comment{levels to consider}
        \Require $C_1,\dots,C_L > 0 $ \Comment{the cost of evaluating $Y_1,\dots,Y_L$ respectively}
        \Require $\btheta_1,\dots,\btheta_L$ \Comment{GP hyperparameters for each level} 
        \Require $n_1,\dots,n_L > 0$ \Comment{the current number of samples on each level}
        \State Set unique $l_1,\dots,l_\tL \in \calL_\mathrm{feasible}$ so that $n_{l_1} C_{l_1} \geq \cdots \geq n_{l_\tL} C_{l_\tL}$ 
        \newline \Comment{order feasible levels by non-increasing cost for doubling the sample size on each level}
        \State $\ell \gets l_1$ \Comment{initialize the selected level}
        \For{$k=2,\dots,\tL$}
            \State $\ell' \gets \ell_k$ \Comment{will compare levels $\ell$ and $\ell'$ where $n_\ell C_\ell \geq n_{\ell'} C_{\ell'}$}
            \State $\hn_{\ell'} \gets n_\ell C_\ell/C_{\ell'} + n_{\ell'}$ \Comment{$n_\ell C_\ell = C_{\ell'} (\hn_{\ell'} - n_{\ell'})$ gives equivalent costs for increasing sample sizes}
            \State $p \gets \lfloor \log_2(\hn_{\ell'}) \rfloor$ \Comment{implies $2^p \leq \hn_{\ell'}  < 2^{p+1}$}
            \State $\hV_{\ell',\hn_{\ell'}} \gets (\hV_{\ell',2^p}^{p+1}/\hV_{\ell',2^{p+1}}^p) \hn_{\ell'}^{\log_2\left(\hV_{\ell',2^{p+1}}/\hV_{\ell',2^p}\right)}$ \Comment{log-log interpolation as in \eqref{eq:logloglininterp}} 
            \State  \Comment{if $n_\ell C_\ell = n_{\ell'} C_{\ell'}$ then one may directly compute $\hV_{\ell',\hn_{\ell'}} = \hV_{\ell',2n_{\ell'}}$ and avoid evaluating $\hV_{\ell',4n_{\ell'}}$}
            \If{$\hV_{\ell',n_{\ell'}} - \hV_{\ell',\hn_{\ell'}} \geq \hV_{\ell,n_\ell} - \hV_{\ell,2n_\ell}$} \Comment{a greater projected decrease in variance for the same cost}
                \State $\ell \gets \ell'$ \Comment{update the selected level}
            \EndIf
        \EndFor
        \\ \Return $\ell$ \Comment{the selected level}
    \end{algorithmic}
\end{algorithm}

\begin{algorithm}[!ht]
    \caption{\texttt{BQMC}: Fast Bayesian Multilevel Quasi-Monte Carlo Without Replications}
    \label{alg:bmlqmc}
    \begin{algorithmic}
        \Require $N>0$ \Comment{the budget}
        \Require $C_1,\dots,C_L > 0 $ \Comment{the cost of evaluating $Y_1,\dots,Y_L$ respectively}
        \Require $n_1^\mathrm{next},\dots,n_L^\mathrm{next} \in \bbN$ powers of two satisfying $\sum_{\ell=1}^L n_\ell^\mathrm{next} C_\ell \leq N$ \Comment{the initial sample sizes} 
        \Require A generating vector $\bg \in \bbN^d$ to use lattices from \Cref{def:lattices} or \\ generating matrices $\mG \in [0,1)^{d \times \infty}$ to use base $2$ digital nets from \Cref{def:dnets}. 
        \Require $\btheta_1,\dots,\btheta_L$ initial hyperparameters containing a global scale $\gamma$, lengthscales $\bEta$, and, if using weighted sums of DSI kernels as in \Cref{def:dsi_kernel}, the kernel weights $\bbeta$ 
        \Require $\bDelta_1,\dots,\bDelta_L \in [0,1)^d$ \Comment{shifts for lattices or digital nets, we use $\bDelta_1,\dots,\bDelta_L \simiid \calU[0,1)^d$.}
        \State $n_\ell \gets 0$ for $\ell \in \{1,\dots,L\}$ \Comment{the number of evaluations}
        \State $\calL \gets \{1,\dots,L\}$ \Comment{the set of levels to update}
        \While{true}
            \State Generate $x_{\ell,i} = \bz_i \oplus \bDelta_\ell$ for $\ell \in \calL$ and $n_\ell \leq i < n_\ell^\mathrm{next}$ \Comment{\Cref{def:lattices,def:dnets}}
            \State Evaluate $Y_\ell(\bx_{\ell,i})$ for $\ell \in \calL$ and $n_\ell \leq i < n_\ell^\mathrm{next}$
            \State Update $\btheta_\ell$ to optimize the NMLL for $\ell \in \calL$
            \State $\hmu_\ell \gets \frac{1}{n_\ell} \sum_{i=0}^{n_\ell-1} Y_\ell(\bx_{\ell,i})$ for $\ell \in \calL$ \Comment{posterior cubature mean}
            \State $\hV_{\ell,n_\ell} \gets \gamma_\ell \left[1-\left(\frac{1}{n_\ell} \sum_{i=0}^{n_\ell-1} \calK_\ell(\bx_{\ell,i},\bx_{\ell,0})\right)^{-1}\right]$ for $\ell \in \calL$ \Comment{posterior cubature variance}
            \State $\calL_\mathrm{feasible} \gets \{\ell \in \{1,\dots,L\}: \sum_{\ell'=1}^L C_{\ell'} n_{\ell'} + C_\ell n_\ell \leq N\}$ \Comment{feasible set of levels}
            \If{$\calL_\mathrm{feasible} = \emptyset$} break \EndIf \Comment{exit while loop if it is not within budget to double on any level}
            \State $\starhat{\ell} \gets \texttt{level\_select\_BQMC}\left(\calL_\mathrm{feasible},(C_\ell,\btheta_\ell,n_\ell)_{\ell=1}^L\right)$ \Comment{choose the level using \Cref{alg:level_select_bmlqmc}}
            \State $\calL \gets \{\starhat{\ell}\}$ and $n_{\starhat{\ell}} \gets n_{\starhat{\ell}}^\mathrm{next}$ and $n_{\starhat{\ell}}^\mathrm{next} \gets 2n_{\starhat{\ell}}$ \Comment{double the sample size on the chosen level}
        \EndWhile
        \State $\hnu \gets \sum_{\ell=1}^L \hmu_\ell$ \Comment{estimate for $\nu$}
        \State $\hV_{n_1,\dots,n_L} \gets \sum_{\ell=1}^L \hV_{\ell,n_\ell}$ \Comment{posterior cubature variance assuming independent GPs $Y_1,\dots,Y_L$}    
        \\ \Return $\hnu,\sqrt{\hV_{n_1,\dots,n_L}},\{n_\ell\}_{\ell=1}^L$ \Comment{the estimate, its standard error, and number of samples per level}
    \end{algorithmic}
\end{algorithm}

\section{Numerical Experiments} \label{sec:numerical_experiments}

In this section we present a number of numerical experiments for both single-level (Q)MC (\Cref{sec:examples_single_level}) and multilevel (Q)MC (\Cref{sec:examples_multilevel}) using the three presented (Q)MC algorithms:
\begin{description}
    \item[MC] (Multilevel) Monte Carlo with IID points as described in \Cref{sec:mlmc,} and \Cref{alg:mlmc}.
    \item[RQMC] (Multilevel) QMC with replications as described in \Cref{sec:mlqmc} and \Cref{alg:mlqmc}.
    \item[BQMC] (Multilevel) Bayesian QMC without replications as described in \Cref{sec:bmlqmc} and \Cref{alg:bmlqmc}.
\end{description}
Our Python implementation relies on the fast Gaussian process regression package \texttt{fastgps} (\url{https://alegresor.github.io/fastgps/}) \citep{sorokin.fastgps_probnum25} and the QMC software package \texttt{qmcpy} (\url{https://qmcsoftware.github.io/QMCSoftware/}) \citep{choi.QMC_software,sorokin.MC_vector_functions_integrals,choi.challenges_great_qmc_software,hickernell.qmc_what_why_how}. We will run $250$ independent trials of each (multilevel) (Q)MC approximation for a given problem and budget. For QMC methods, we will consider both the lattices in \Cref{def:lattices} and the digital nets in \Cref{def:dnets}. For lattices, we will use the \texttt{kuo.lattice-33002-1024-1048576.9125} generating vector from \url{https://web.maths.unsw.edu.au/~fkuo/lattice/} \citep{cools2006constructing,nuyens2006fast}, and for digital nets, we will use the \texttt{joe-kuo-6.21201} generating matrices from \url{https://web.maths.unsw.edu.au/~fkuo/sobol/index.html} \citep{joe2003remark,joe2008constructing}. These are the default choices in \texttt{qmcpy}, which sources copies of these generating vectors and matrices in standardized formats from the \texttt{LDData} repository \url{https://github.com/QMCSoftware/LDData/}. Comparisons against other choices of generating vectors and matrices is a valuable avenue for future work. 
To simplify presentation, we will only present results for digital nets in the main text with analogous results for lattices deferred to \Cref{sec:appendixA}. 

Most of the single-level test functions in \Cref{sec:examples_single_level} were considered by \citet{lecuyer.RQMC_CLT_bootstrap_comparison}, albeit with the ridge functions having equal weights. There, comprehensive experiments showed that RQMC methods outperform alternative bootstrap methods in terms of confidence interval coverage across a range of benchmarks. Their experiments tested up to dimension $d=32$, and considered all randomizations options $R \in \{5,10,20,30\}$. They reported RQMC coverage failures only for $R=5$ randomizations. We will use $R=8$ randomizations in our RQMC testing. As $R$ decreases, RQMC will typically yield better true errors with less accurate standard errors. Our choice of $R=8$ is rather aggressive in comparison to existing literature which often uses at least $R=25$.

\subsection{Single Level Problems} \label{sec:examples_single_level}

Here we consider single level test functions $Y$ and run (Q)MC algorithms to estimate $\mu = \bbE[Y(\bX)]$ for $\bX \sim \calU[0,1]^d$ and dimension $d=32$. We consider the four test functions listed below. The two ridge functions are defined as $Y(\bx) = g(u(\bx))$ for $u(\bx) = \sum_{j=1}^d c_j \Phi^{-1}(x_j)$ with weights $(c_j)_{j=1}^d$ and $\Phi$ the CDF of the standard normal $\calN(0,1)$. We only present results for \emph{sparse weights} $c_j = 2^{-j} / \sqrt{\sum_{j'=1}^d 2^{-2j'}}$ which make the problem QMC-friendly, but we observed similar findings for ridge functions with \emph{equal weights} $c_j = d^{-1/2}$.  Both sparse and equal weights make $u(\bX) \sim \calN(0,1)$ when $\bX \sim \calU[0,1]^d$.

\begin{description}
    \item[sumxex] $Y(\bx) = -d+\sum_{j=1}^d x_j\exp(x_j)$. This is a smooth and additive integrand which is easy for QMC methods.
    \item[ridge PL sparse] $Y(\bx) = g(u(\bx))$ with $g(u)=\max\left(u-1,0\right) - \phi(1)+\Phi(-1)$ and $\phi$ the density of $\calN(0,1)$. This is a continuous piecewise linear (PL) function with a kink, a feature often observed in problems from computational finance. 
    \item[ridge JSU sparse] $Y(\bx) = g(u(\bx))$ with $g(u) = -\eta+F^{-1}\left(\Phi\left(v\right)\right)$ where $F$ is the CDF of a Johnson's SU distribution \citep{johnson1949systems} with parameters $\gamma=\delta=\lambda=1$ and $\xi=0$, and $\eta$ is the mean of that distribution. For $\bX \sim \calU[0,1]^d$, $Y(\bX)$ has skewness $-5.66$ and kurtosis $96.8$ making it heavy tailed. 
    \item[Genz corner-peak 2] $Y(\bx) = \left(1+\sum_{j=1}^d c_j x_j\right)^{-(d+1)}$ with coefficients $c_j = j^{-2}/(4\sum_{j'=1}^d (j')^{-2})$. This is a corner-peak version of the integrand in \citet{genz1993comparison} (as opposed to the oscillatory version) with coefficients of the second kind (out of three common options). The different coefficient options represent increasing levels of anisotropy and decreasing effective dimension.
\end{description}

\begin{figure}[t]
    \centering
    \includegraphics[width=1\textwidth]{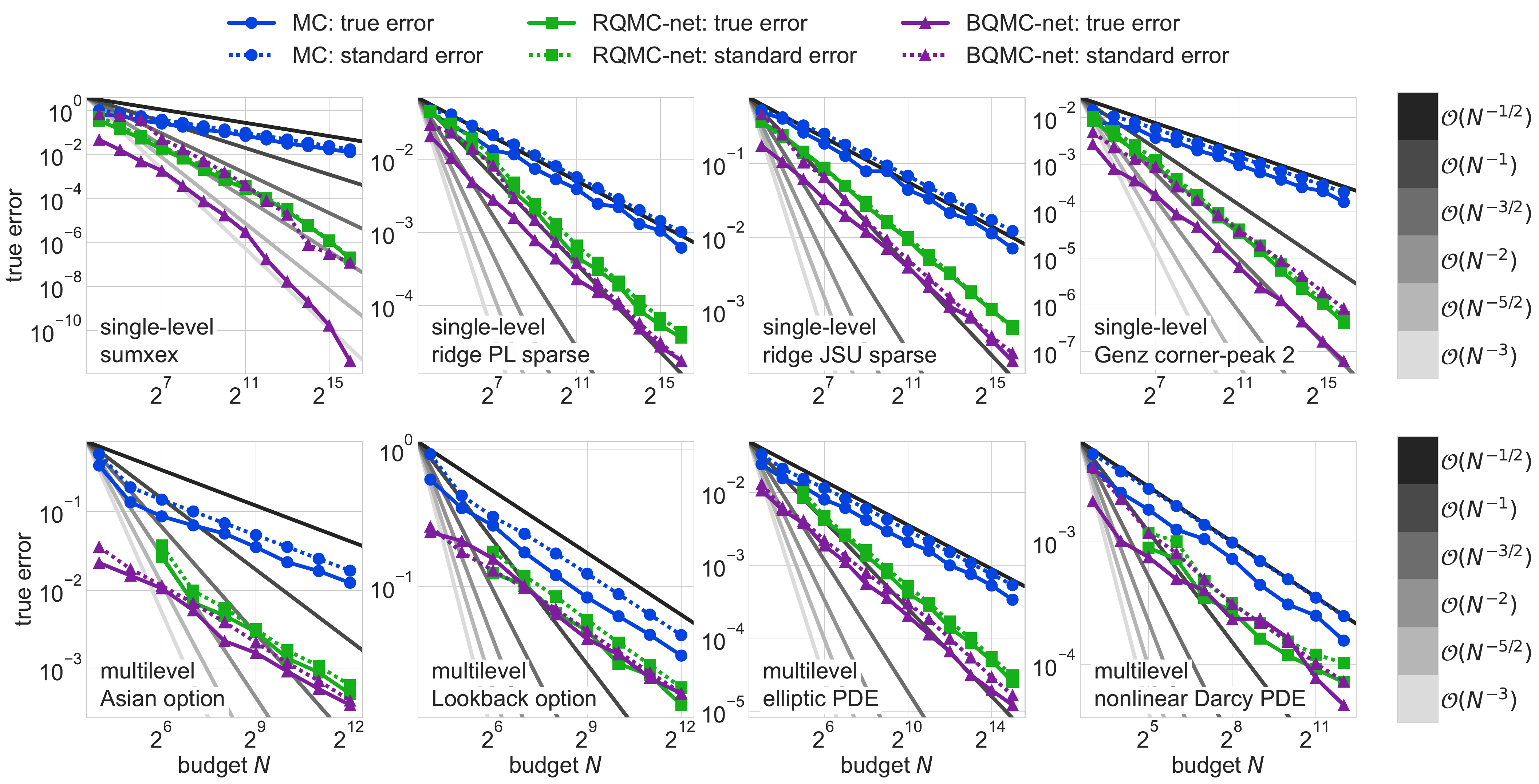}
    \caption{Median true error and standard error versus budget for Monte Carlo (MC) with IID points, quasi-Monte Carlo with replications (RQMC), and quasi-Monte Carlo with fast Bayesian cubature (BQMC). Here QMC methods use digital nets, see \Cref{fig:lattice.convergence} for the lattice version.} 
    \label{fig:dnet.convergence}
\end{figure}

In the top row of \Cref{fig:dnet.convergence}, we plot both the median true errors and median standard errors for single-level problems with an increasing budget 
using digital nets. For these single-level problems, the true error and standard error for IID Monte Carlo always converge at the expected $\calO(n^{-1/2})$ rate with the true errors closely matching the standard errors. The RQMC methods also yield standard errors which closely match true errors, with both converging faster than IID MC methods. The RQMC method converges like $\calO(n^{-2})$ for the easy sumxex problem, slightly below $\calO(n^{-1})$ for both the ridge PL sparse and ridge JSU sparse problems, and between $\calO(n^{-1})$ and $\calO(n^{-3/2})$ for the Genz corner-peak 2 function. As expected, the BQMC methods almost always provide better true errors compared to the RQMC methods. In the majority of cases, the true error rates for BQMC match those for the RQMC counterparts, but with smaller leading constants. One exception is the true errors of BQMC for the sumxex problem which appears to converge at a rate slightly better than $\calO(n^{-3})$ compared to the $\calO(n^{-2})$ rate for the RQMC method.

We found the BQMC standard errors are often able to match the true error convergence rates and provide slight improvements over the corresponding RQMC standard errors. However, the rate constants for the BQMC standard errors are rather conservative for their true errors when higher-order convergence beyond $\calO(n^{-2})$ is observed. For example, when applying BQMC to the easy sumxex problem, the true error converges like $\calO(n^{-3})$ while the standard error converges with a rate around $\calO(n^{-2})$. One potential method to improve BQMC standard error convergence would be to expand the DSI sum kernels in \Cref{def:dsi_kernel} to consider even higher-order smoothness kernels beyond $\alpha=4$. However, evaluating higher-order smoothness kernels is more expensive in terms of both computations and memory, and a larger set of hyperparameters weights $\bbeta$ would need to be optimized. We also found the BQMC standard error convergence rate appears slightly worse than that for RQMC for the Genz corner-peak $2$ function. 

\begin{figure}[!ht]
    \centering
    \includegraphics[width=1\textwidth]{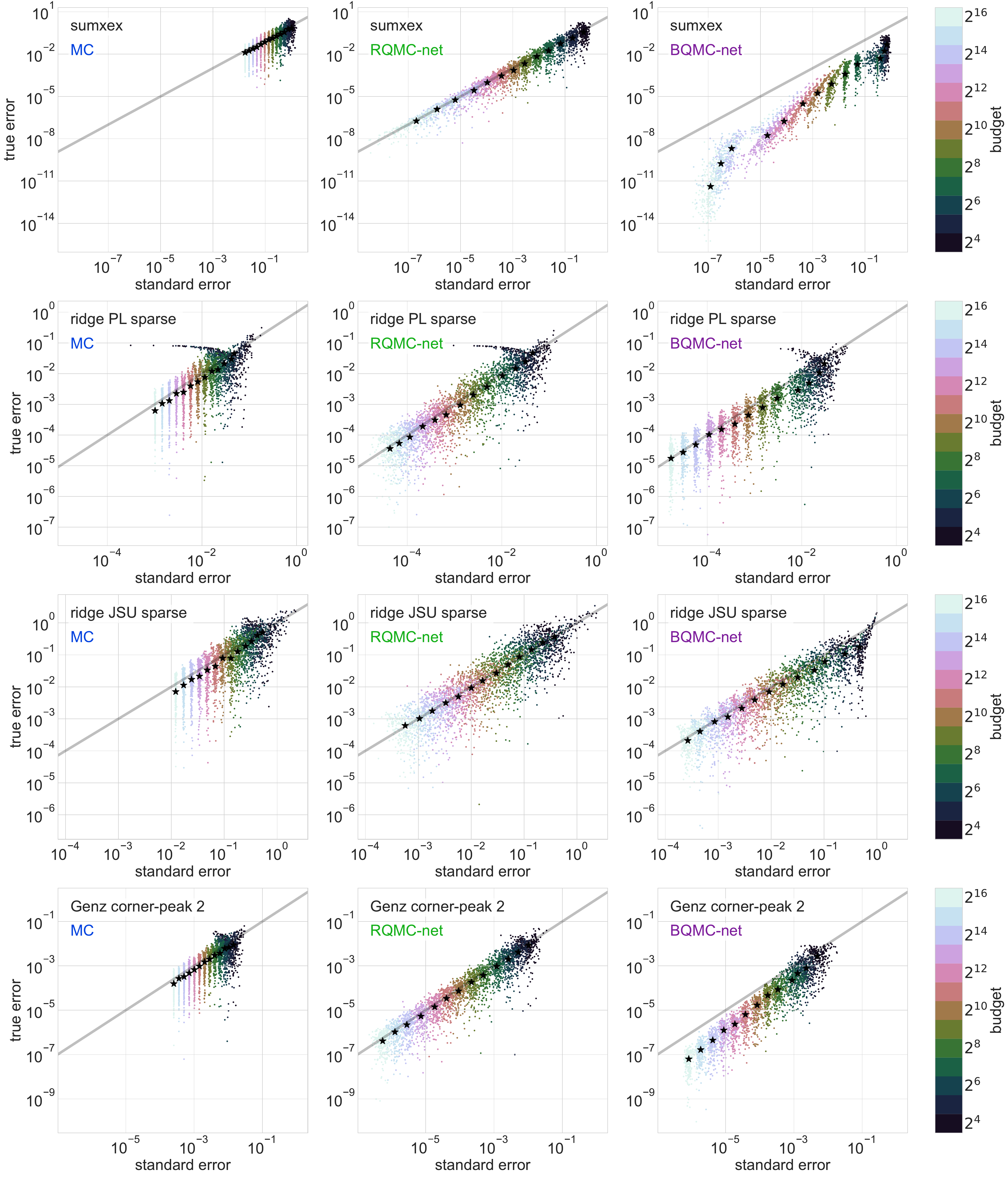}
    \caption{Standard error versus true error across trials for (single level) Monte Carlo (MC) with IID points, quasi-Monte Carlo with replications (RQMC), and quasi-Monte Carlo with fast Bayesian cubature (BQMC). The stars represent the median true and standard errors for each budget. Here QMC methods use digital nets, see \Cref{fig:lattice.sl.error_scatters} for the lattice version.} 
    \label{fig:dnet.sl.error_scatters}
\end{figure}

In \Cref{fig:dnet.sl.error_scatters} we plot the standard error and true error for each trial against each other for digital nets. Again, we see the conservative nature of the BQMC standard error estimations in comparison to RQMC methods. This is especially evident for the sumxex and Genz corner-peak 2 functions at higher budgets $n$. For the two ridge functions, we found the BQMC standard errors were accurate with respect to the true errors, with close agreement in both medians.

\subsection{Multilevel Problems} \label{sec:examples_multilevel}

Here we consider multilevel test functions with a fixed number of levels. In the following subsections we will describe our test functions for option pricing (\Cref{sec:option_pricing}), solving a one-dimensional elliptic PDE (\Cref{sec:elliptic_pde}), and solving a two-dimensional nonlinear Darcy flow PDE (\Cref{sec:darcy}). Our multilevel numerics follow in \Cref{sec:numerics_ml}.  \Cref{tab:ml_decay} gives approximate means $\mu_\ell$ and standard deviations $\sqrt{V_\ell}$ for each problem. Our cost always increases with level, and we always normalize so that the cost on the maximum level is $1$.  

\begin{table}[!ht]
    \centering
    \begin{tabular}{rrrrrrrrr} 
        \multicolumn{9}{c}{mean $\mu_\ell = \bbE[Y_\ell(\bX)] = \bbE[Q_\ell(\bX)-Q_{\ell-1}(\bX)]$ with $\bX \sim \calU[0,1]^d$} \\
        problem/$\ell$ & 1 & 2 & 3 & 4 & 5 & 6 & 7 & 8 \\
        \hline 
        Asian option & 6.3e0  & -3.0e-1 & -1.5e-1 & -7.2e-2 & -3.7e-2 & -1.8e-2 & -9.3e-3 & -4.8e-3 \\
        Lookback option & 1.3e1 & 1.4e0 & 8.9e-1 & 5.8e-1& 4.1e-1 & 2.8e-1 & 1.8e-1 & 1.3e-1 \\
        elliptic PDE & 1.6e-1 & -1.1e-2 & 2.5e-3 & 1.5e-3  \\
        nonlinear Darcy PDE & 4.1e-2 & 8.5e-4 & 3.3e-4 \\
        \hline 
        \hline 
        \multicolumn{9}{c}{standard deviation $\sqrt{V_\ell} = \sqrt{\bbV[Y_\ell(\bX)]} = \sqrt{\bbV[Q_\ell(\bX)-Q_{\ell-1}(\bX)]}$ with $\bX \sim \calU[0,1]^d$} \\
        problem/$\ell$ & 1 & 2 & 3 & 4 & 5 & 6 & 7 & 8 \\
        \hline 
        Asian option & 8.7e0 & 6.8e-1 & 3.5e-1 & 1.7e-1 & 8.4e-2 & 4.3e-2 & 2.1e-2 & 1.1e-2 \\
        Lookback option & 1.3e1 & 2.1e0 & 1.4e0 & 9.0e-1 & 6.3e-1 & 4.2e-1 & 2.9e-1 & 2.0e-1 \\
        elliptic PDE & 1.4e-1 & 6.2e-2 & 1.0e-2 & 3.5e-3 \\
        nonlinear Darcy PDE & 7.8e-2 & 6.7e-3 & 1.9e-3
    \end{tabular}
    \caption{Decay of the means $\mu_\ell$ and standard deviations $\sqrt{V_\ell}$ of differences with increasing level $\ell$.}
    \label{tab:ml_decay}
\end{table}

\subsubsection{Option Pricing} \label{sec:option_pricing}

Suppose we are given a financial option with starting price $S_0$, strike price $K$, interest rate $r$, and volatility $\sigma$ which is exercised at time $1$. We use $S_0=K=100$, $r=0.05$, and $\sigma=0.2$. We will assume the asset path follows a geometric Brownian motion $S(t) = S_0 e^{(r-\sigma^2/2)t+\sigma B(t)}$ where $B(t)$ is a standard Brownian motion. At level $\ell$, we monitor the asset at $d_\ell = 2^{2+\ell}$ times $(j/d_\ell)_{j=1}^{d_\ell}$ so that $(B(1/d_\ell),B(2/d_\ell),\dots,B(1))^\intercal \sim \calN\left(\bzero,\mSigma_\ell\right)$ where $\mSigma_\ell = \left(\min(j/d_\ell,j'/d_\ell)\right)_{j,j'=1}^{d_\ell}$. We will use the eigendecomposition $\mSigma_\ell = \mA_\ell \mA_\ell^\intercal$ to write $(B(1/d_\ell),B(2/d_\ell),\dots,B(1))^\intercal \sim \mA_\ell \Phi^{-1}(\bX_\ell)$ where $\Phi^{-1}$ is the inverse CDF of the standard normal applied elementwise to $\bX_\ell \sim \calU[0,1]^{d_\ell}$. The cost $C_\ell$ on level $\ell$ is proportional to $2^\ell$. We will consider the following two options, both with $L=8$ levels of discretization.

\begin{description}
    \item[Asian Call Option with Geometric Averaging] The discounted fair price of the infinite-dimensional option with continuous monitoring is $\bbE\left[\max\left(\exp\left(\int_0^1 \log(S(t)) \D t\right) - K,0\right)\right]e^{-r}$, and for discrete monitoring at level $\ell$ we use the finite-dimensional approximation 
    $$Q_\ell(\bX) = \max\left(\prod_{j=1}^{d_\ell} S(j/d_\ell) - K,0\right)e^{-r}.$$ 
    \item[Lookback Option] The discounted fair price of the infinite-dimensional option with continuous monitoring is $\bbE\left[S(1)-\min_{0 \leq t \leq 1} S(t)\right] e^{-r}$, and for discrete monitoring at level $\ell$ we use the finite-dimensional approximation 
    $$Q_\ell(\bX) = \left[S(1) - \min(S(1/d_\ell),S(2/d_\ell),\dots,S(1))\right] e^{-r}.$$
\end{description}

\subsubsection{Elliptic PDE} \label{sec:elliptic_pde}

Let us consider the one-dimensional elliptic PDE $-\nabla(e^{a(u,\bx)} \nabla q(u,\bx)) = g(u)$ with $u \in [0,1]$ and boundary conditions $q(0,\bx) = q(1,\bx) = 0$ for all $\bx \in [0,1]^d$. The forcing term $g$ is set to the constant $g(u) = 1$ for all $u \in [0,1]$. We will generate $a(u,\bX) = \sum_{j=1}^d \Phi^{-1}(X_j) \sin(\pi k u)/j$ with $\bX \sim \calU[0,1]^d$ in $d=8$ dimensions and $\Phi$ the CDF of the standard normal so $\Phi^{-1}(X_j) \sim \calN(0,1)$. Let us denote by $q_\ell$ the level $\ell$ numerical PDE solution using a finite difference method with $2^{1+\ell}+1$ evenly spaced mesh points $(k/2^{1+\ell})_{k=0}^{2^{1+\ell}}$. We will take the discretized solution to be $Q_\ell(\bX) = q_\ell(1/2,\bX)$ and use $L=4$ levels. For each query of $Q_\ell$ we need to solve a tridiagonal linear system, which can be done with linear complexity. Therefore, we take the cost $C_\ell$ on level $\ell$ to be proportional to $2^\ell$. 

\subsubsection{Nonlinear Darcy Flow PDE} \label{sec:darcy}

\begin{figure}[!ht]
    \centering
    \includegraphics[width=1\textwidth]{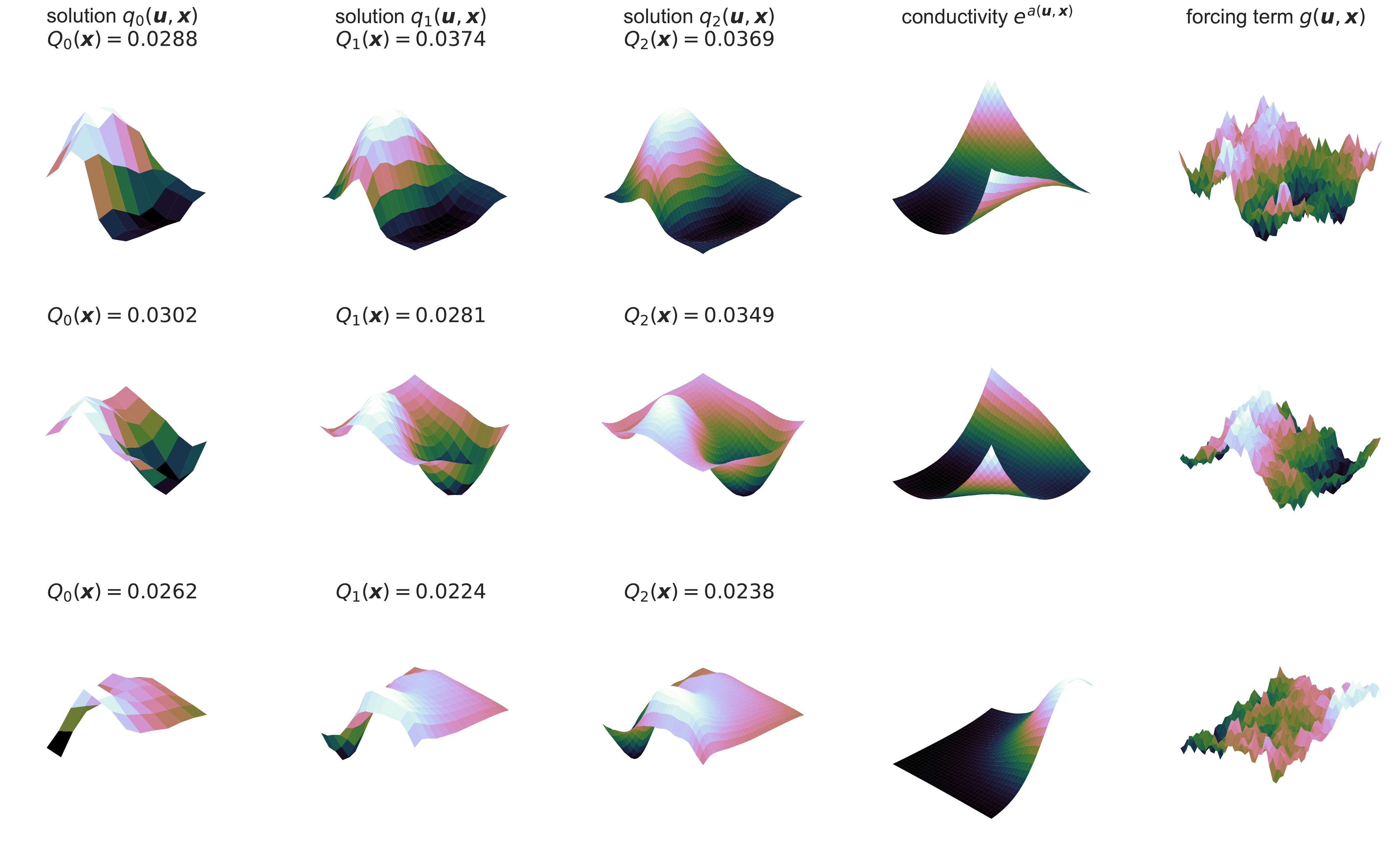}
    \caption{Visualization of the nonlinear Darcy flow PDE where each row has a different realization $\bx$ of both the random conductivity and forcing term. The resolution increases with each of the $L=3$ levels, with the conductivities and forcing terms shown at their maximum resolutions. Here the quantity of interest on level $\ell\in \{1,2,3\}$ is $Q_\ell(\bX) = \max_{\bu \in [0,1]^d} q_\ell(\bu,\bX)$ for $\bX \sim \calU[0,1]^d$. See \Cref{sec:darcy} for details.} 
    \label{fig:darcy}
\end{figure}

We now consider a two-dimensional nonlinear Darcy flow problem 
\begin{equation*} \label{eq:Darcy}
\begin{cases}
    - \nabla \cdot (e^{a(\bu,\bx)} . \nabla q(\bu,\bx)) +  (q(\bu,\bx))^3 = g(\bu,\bx), & \bu \in [0,1]^2 \\
    q(\bu,\bx) = 0, & \bu \in \partial [0,1]^2
\end{cases}
\end{equation*}
where $g$ is a forcing term and $e^a$ represents the conductivity. Here we assume $g$ and $a$ are both random and independently drawn from Gaussian processes with the underlying stochasticity of both processes controlled by $\bX \sim \calU[0,1]^d$. We will take $g$ to be a draw from a GP with a $1/2$-Mat\'ern kernel and $a$ to be a draw from a GP with a Gaussian kernel. 
As with the elliptic PDE example in \Cref{sec:elliptic_pde}, we will let $q_\ell$ denote the level $\ell$ numerical PDE solution solved on a computational grid with $2^{2+\ell}-1$ mesh points $(k/2^{2+\ell})_{k=1}^{2^{2+\ell}-1}$ in each dimension. We numerically solve the PDE via iterative linearizations and a Levenberg--Marquardt scheme \citep{levenberg1944method,marquardt1963algorithm}. Our quantity of interest is $\bbE[\max_\bu(q(\bu,\bx))]$, so we take the discretized solution to be $Q_\ell(\bX) = \max_\bu q_\ell(\bu,\bX)$ with $\bX \sim \calU[0,1]^d$. \Cref{fig:darcy} visualizes realizations of the conductivity, forcing term, and PDE solution at varying levels of resolution. Theoretically, the cost of evaluating $Q_\ell$ is $\calO(S_\ell 2^{6\ell})$, with $S_\ell$ the number of Levenberg--Marquardt steps, as at each step we need to solve a dense linear system in a $(2^{2+2\ell}-1) \times (2^{2+2\ell}-1)$ matrix. Practically, we take $L=3$ levels and set the costs proportional to the respective average runtimes in seconds $(4.2 \times 10^{-5},9.9 \times 10^{-5},2.8 \times 10^{-3})$. Due to the large scale nature of this problem with $d=3844$ dimensions, we chose to only consider the $\alpha=4$ DSI kernel in \Cref{def:dsi_kernel}, i.e., we fixed $\beta_1=\beta_2=\beta_3=0$ and $\beta_4=1$. This enabled faster hyperparameter optimization and reduced storage. Even so, our Darcy flow experiment required over $12$ hours running in parallel on $5$ NVIDIA A100 80 GB GPUs. 

\subsubsection{Results for Multilevel Numerical Experiments} \label{sec:numerics_ml}

In the bottom row of \Cref{fig:dnet.convergence} we plot both the median true errors and median standard errors for multilevel problems with an increasing budget $N$ using digital nets. As was the case with the single-level problems, for these multilevel problems we again find IID Monte Carlo methods achieve matching true and standard errors with convergence like $\calO(N^{-1/2})$ for each problem. The RQMC methods also yielded accurate standard error estimates which closely matched the true errors for all problems. RQMC converged like $\calO(N^{-1/2})$ for both option pricing problems and the Darcy flow PDE with smaller rate constants than IID Monte Carlo in each case. For the elliptic PDE, RQMC converges like $\calO(N^{-1})$. For BQMC, we are able to accommodate smaller budgets $N$ compared to the RQMC methods which require multiple replications, as evidenced by the curves extending further to the left for BQMC than RQMC.

As with single level problems, BQMC outperforms RQMC in terms of both true and standard errors in the vast majority of cases. We observed a better rate constant for BQMC than RQMC on both the Asian option and elliptic PDE examples. On the harder Lookback option and nonlinear Darcy examples, we observed similar errors between BQMC and RQMC. We did observe slightly better RQMC median true errors for the Darcy problem with the $N=2^9$ and $N=2^{10}$ budgets compared to BQMC, but BQMC then yields better true errors at our largest tested budgets of $N=2^{11}$ and $N=2^{12}$. Due to the lack of higher-order convergence, we also found the BQMC standard error estimate to be accurate for the true errors, even in small sample regimes. 

\begin{figure}[!ht]
    \centering
    \includegraphics[width=1\textwidth]{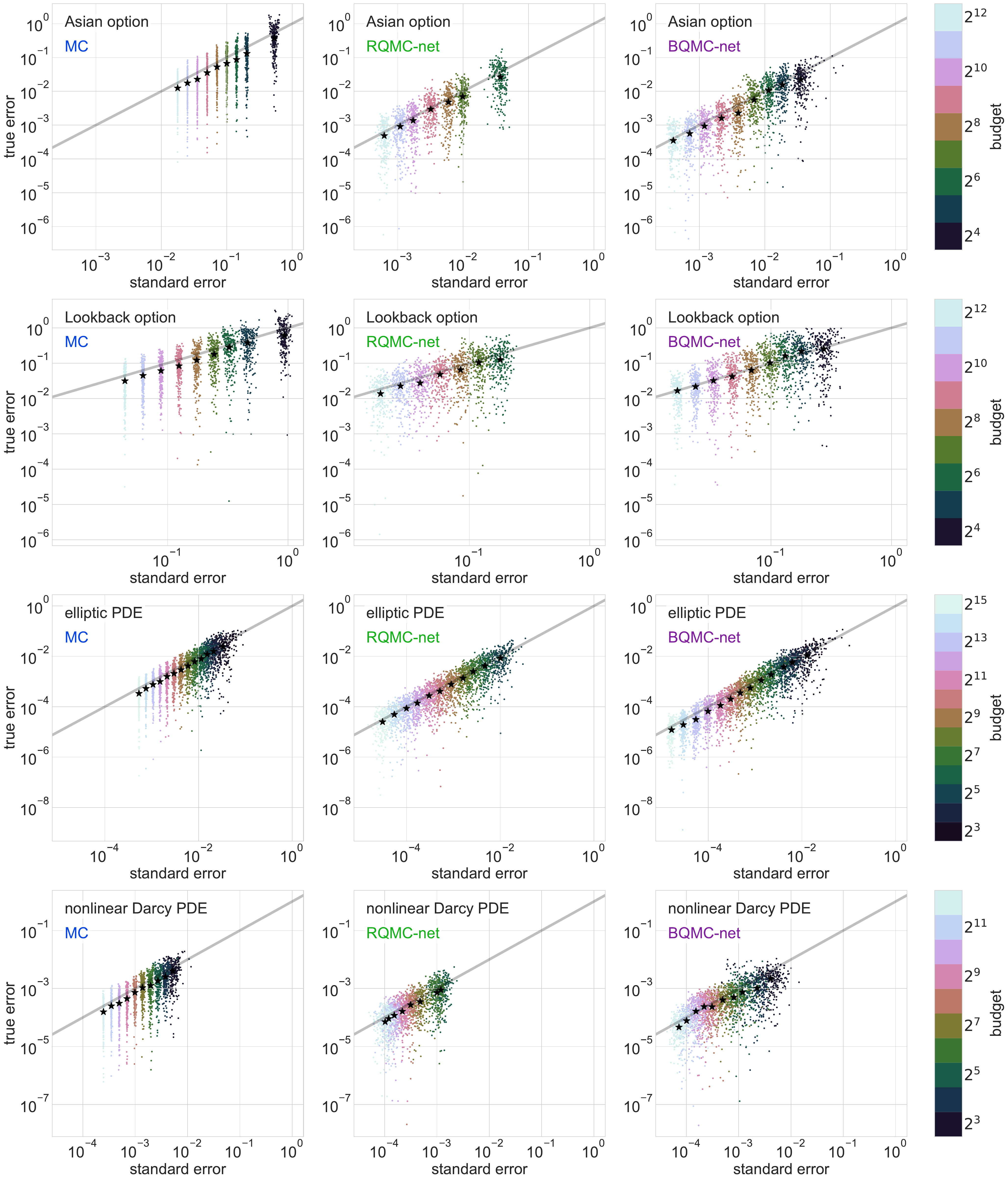}
    \caption{Standard error versus true error across trials for multilevel Monte Carlo (MC) with IID points, quasi-Monte Carlo with replications (RQMC), and quasi-Monte Carlo with fast Bayesian cubature (BQMC). The stars represent the median true and standard errors for each budget. Here QMC methods use digital nets, see \Cref{fig:lattice.ml.error_scatters} for the lattice version.} 
    \label{fig:dnet.ml.error_scatters}
\end{figure}

In \Cref{fig:dnet.ml.error_scatters} we plot the standard error and true error for each trial against each other for digital nets. Here, the accurate standard error estimates from BQMC for the Darcy flow problem and high-budget Asian option trials are more readily evident. Across both BQMC methods, we observed that the standard errors become tighter across the trials as the budget increases. In order words, for larger budgets $N$, the BQMC methods become more consistent and confident in their standard error predictions. This feature is shared with IID Monte Carlo methods, but not with RQMC methods, which always aggregate a fixed number of independent estimates $R$. 

\begin{figure}[!ht]
    \centering
    \includegraphics[width=1\textwidth]{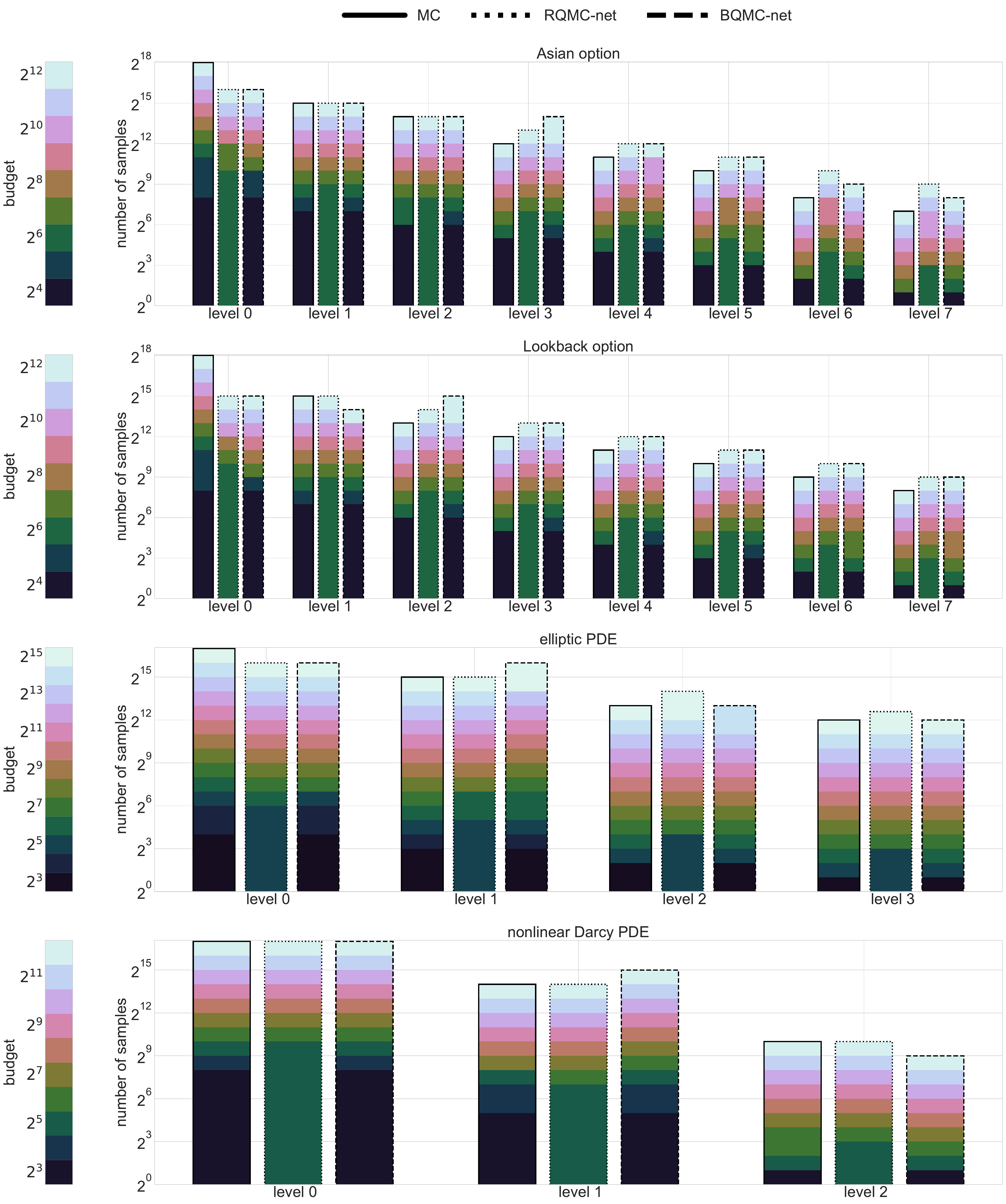}
    \caption{Sample allocation by budget versus level for multilevel Monte Carlo (MC) with IID points, quasi-Monte Carlo with replications (RQMC), and quasi-Monte Carlo with fast Bayesian cubature (BQMC). Here QMC methods use digital nets, see \Cref{fig:lattice.ml.sample_allocation_group_level} for the lattice version.} 
    \label{fig:dnet.ml.sample_allocation_group_level}
\end{figure}

Finally, in \Cref{fig:dnet.ml.sample_allocation_group_level} we plot the sample allocation versus budget for digital nets. IID Monte Carlo often seems to allocate a greater proportion of the budget to lower levels in comparison to RQMC and BQMC methods. The RQMC and BQMC methods yielded similar allocations across most problems for both point sets. BQMC did show a slight preference for allocating more budget to moderate levels while RQMC tended to allocate a slightly greater proportion of the budget to the high and low levels. 

\section{Conclusion} \label{sec:conclusion}

This article has proposed a fast Bayesian MLQMC algorithm which requires only a single low-discrepancy sequence on each level. We performed a number of numerical experiments which show the proposed method is more efficient than the standard MLQMC method which uses multiple replications on each level. In the development of our fast method, we proposed a novel Bayesian sample allocation scheme which exploits the ability of GPs to forecast future error estimates before actually evaluating the function at the future points. We also put forward a new digitally-shift-invariant kernel with adaptive smoothness that displays strong performance on our test problems. 

\section*{Acknowledgements}

This material is based upon work supported by the U.S. Department of Energy, Office of Science, Office of Workforce Development for Teachers and Scientists, Office of Science Graduate Student Research (SCGSR) program. The SCGSR program is administered by the Oak Ridge Institute for Science and Education for the DOE under contract number DE-SC0014664.

This work is supported by National Science Foundation DMS Grant No. 2316011. 

This work was supported by the Laboratory Directed Research and Development (LDRD) program and the Advanced Simulation and Computing, Verification and Validation (ASC V\&V) program at Sandia National Laboratories.

This article has been co-authored by employees of National Technology and Engineering Solutions of Sandia, LLC under Contract No.\,DE-NA0003525 with the U.S. Department of Energy (DOE). The employees co-own right, title and interest in and to the article and are responsible for its contents. The United States Government retains and the publisher, by accepting the article for publication, acknowledges that the United States Government retains a non-exclusive, paid-up, irrevocable, world-wide license to publish or reproduce the published form of this article or allow others to do so, for United States Government purposes. The DOE will provide public access to these results of federally sponsored research in accordance with the DOE Public Access Plan available at \url{https://www.energy.gov/downloads/doe-public-access-plan}.

\bibliography{main}

\appendix

\renewcommand{\thefigure}{A.\arabic{figure}}

\section{Additional numerical results for lattices} \label{sec:appendixA}

For both the RQMC and BQCM lattice methods, we always apply a periodizing tent transform $x \mapsto 1-2 \lvert x-1/2 \rvert$ which preserves continuity but may introduce discontinuities in the derivative of the integrand. Smoothness preserving transforms have been used for fast Bayesian cubature \citep{rathinavel.bayesian_QMC_lattice,rathinavel.bayesian_QMC_thesis}, but they often create regions of higher variation which make QMC methods less effective. A more systematic comparison of smoothness preserving transforms matched with higher-order smoothness SI kernels in \Cref{def:si_kernel} would be a valuable exploration for future work. Here we will use the lowest-smoothness ($\alpha=1$) SI kernel to assume as little periodicity as possible.

\begin{figure}[!ht] 
    \centering
    \includegraphics[width=1\textwidth]{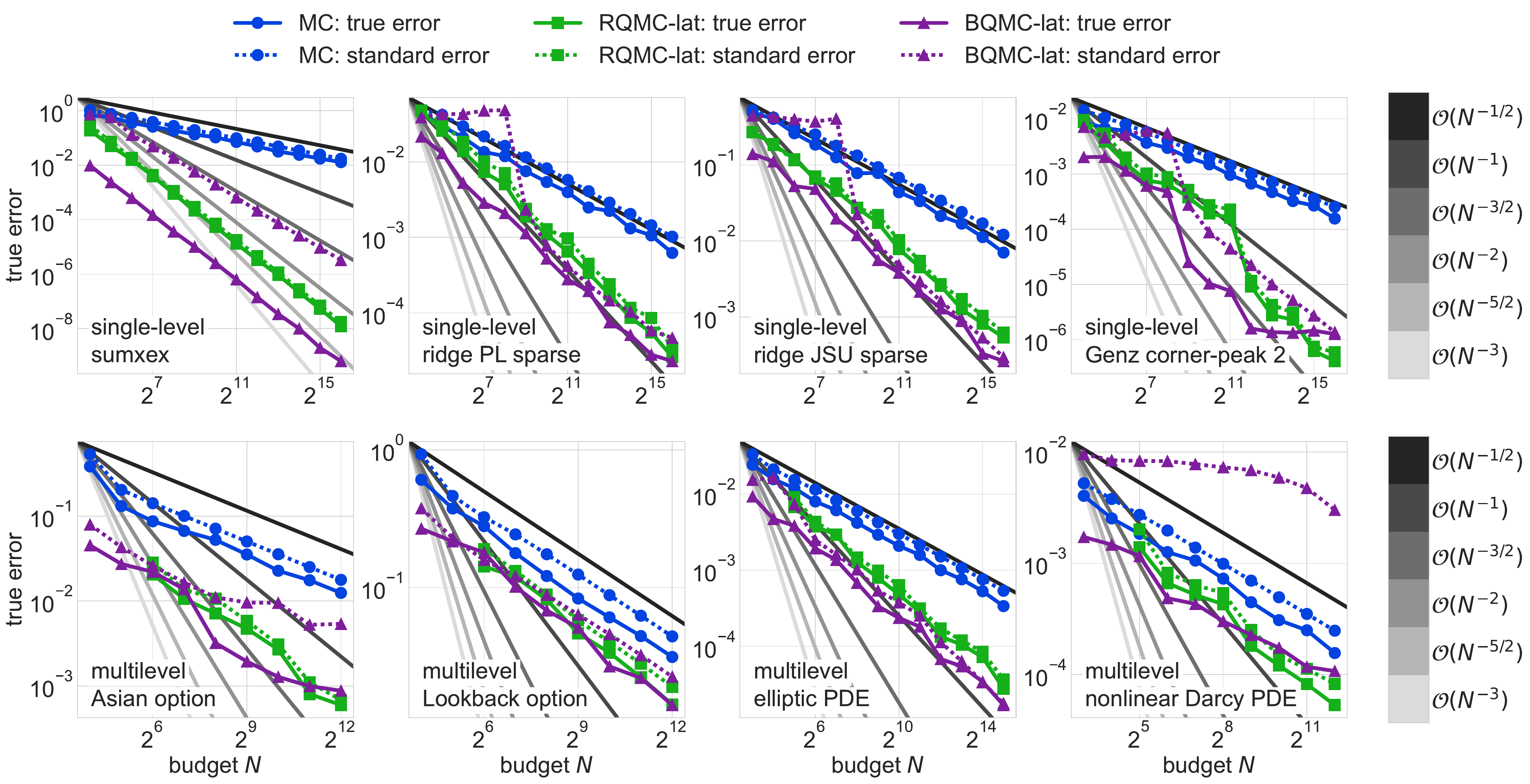}
    \caption{Median true error and standard error versus budget for Monte Carlo (MC) with IID points, quasi-Monte Carlo with replications (RQMC), and quasi-Monte Carlo with fast Bayesian cubature (BQMC). Here QMC methods use lattices, see \Cref{fig:dnet.convergence} for the digital net version.} 
    \label{fig:lattice.convergence}
\end{figure}

\begin{figure}[!ht]
    \centering
    \includegraphics[width=1\textwidth]{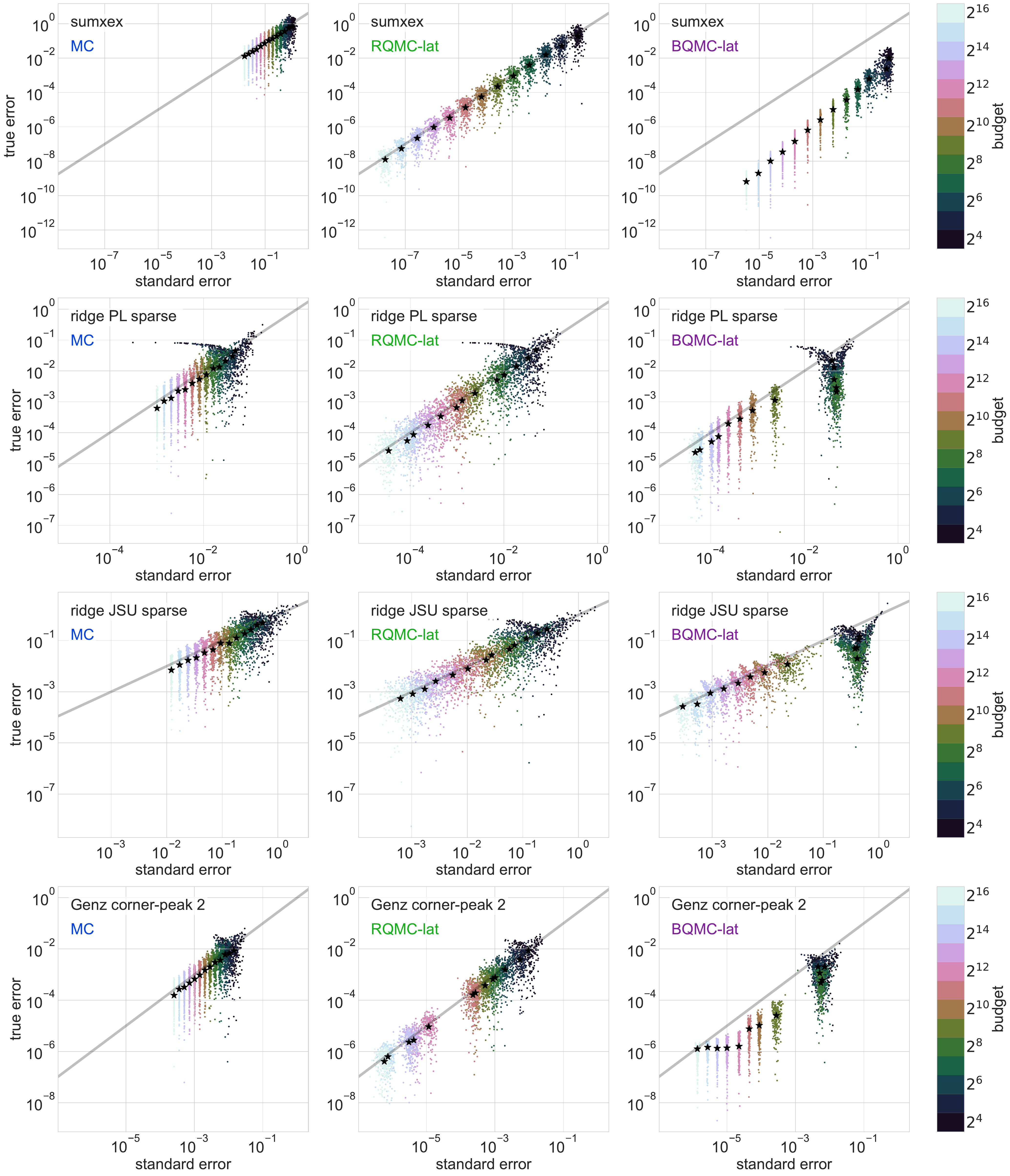}
    \caption{Standard error versus true error across trials for (single level) Monte Carlo (MC) with IID points, quasi-Monte Carlo with replications (RQMC), and quasi-Monte Carlo with fast Bayesian cubature (BQMC). The stars represent the median true and standard errors for each budget. Here QMC methods use lattices, see \Cref{fig:dnet.sl.error_scatters} for the digital net version.} 
    \label{fig:lattice.sl.error_scatters}
\end{figure}

\begin{figure}[!ht]
    \centering
    \includegraphics[width=1\textwidth]{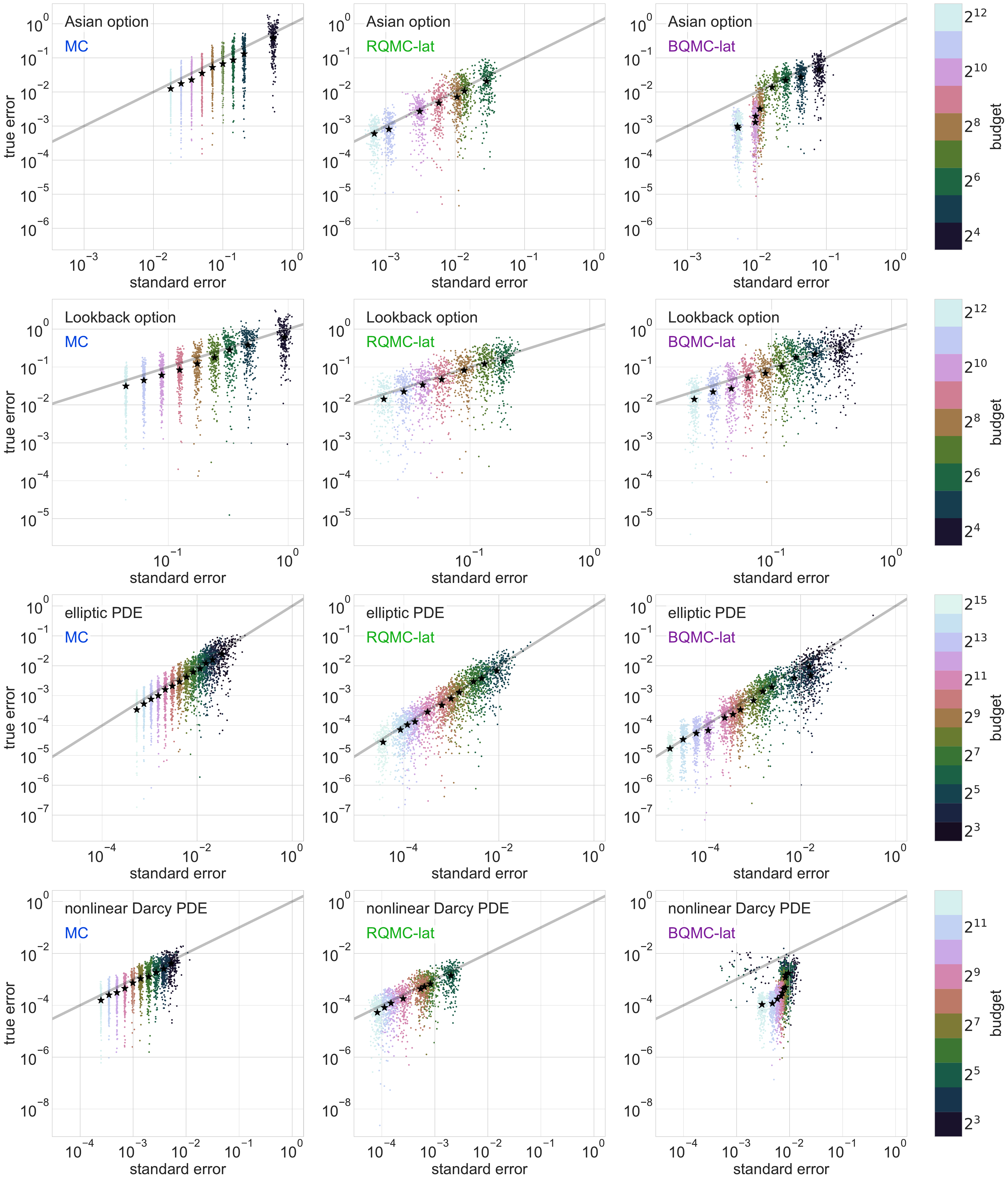}
    \caption{Standard error versus true error across trials for multilevel Monte Carlo (MC) with IID points, quasi-Monte Carlo with replications (RQMC), and quasi-Monte Carlo with fast Bayesian cubature (BQMC). The stars represent the median true and standard errors for each budget. Here QMC methods use lattices, see \Cref{fig:dnet.ml.error_scatters} for the digital net version.} 
    \label{fig:lattice.ml.error_scatters}
\end{figure}

\begin{figure}[!ht]
    \centering
    \includegraphics[width=1\textwidth]{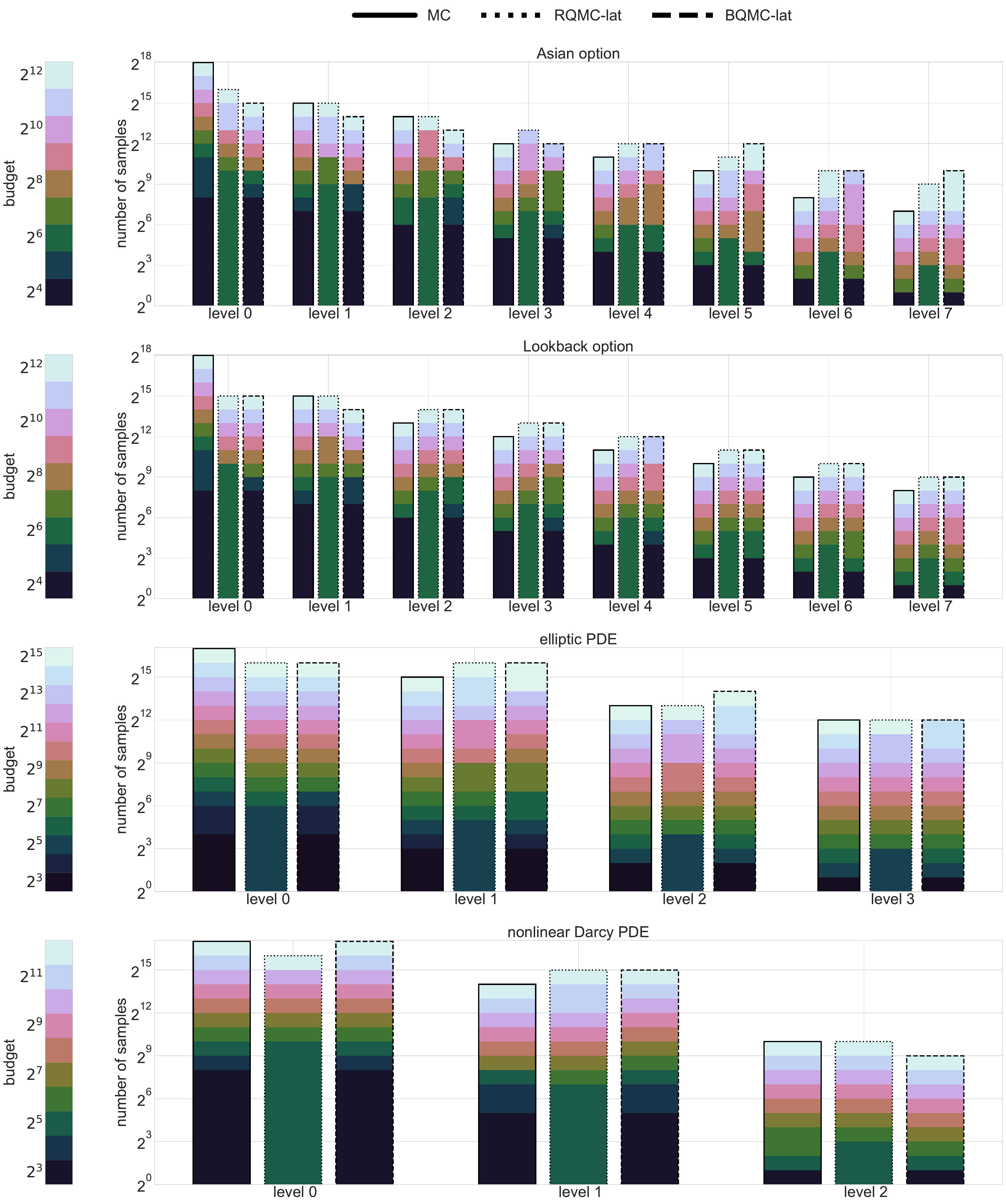}
    \caption{Sample allocation by budget versus level for multilevel Monte Carlo (MC) with IID points, quasi-Monte Carlo with replications (RQMC), and quasi-Monte Carlo with fast Bayesian cubature (BQMC). Here QMC methods use lattices, see \Cref{fig:dnet.ml.sample_allocation_group_level} for the digital net version.} 
    \label{fig:lattice.ml.sample_allocation_group_level}
\end{figure}

\Cref{fig:lattice.convergence} shows the median true errors and median standard errors when using lattice rules (see \Cref{fig:dnet.convergence} for the digital net equivalent). The top row contains results for the single-level problems in \Cref{sec:examples_single_level}, while the bottom row contains results for the multilevel problems in \Cref{sec:examples_multilevel}. \Cref{fig:lattice.sl.error_scatters} and \Cref{fig:lattice.ml.error_scatters} plot standard errors against true errors for each trial using lattices for the single-level and multilevel problems respectively (see \Cref{fig:dnet.sl.error_scatters} and \Cref{fig:dnet.ml.error_scatters} for the respective digital net equivalents). \Cref{fig:lattice.ml.sample_allocation_group_level} shows the sample allocations per level across increasing budgets for the multilevel problems  using lattices (see \Cref{fig:dnet.ml.sample_allocation_group_level} for the digital net equivalent).

For lattices, in the majority of cases the true error rates for BQMC match those for the RQMC counterparts, but with smaller leading constants. The RQMC methods yielded accurate standard error estimates which closely matched the true errors for all problems. For BQMC-lat, we found the standard error for the single-level sumxex problem had a rate of $\calO(n^{-3/2})$, which was not able to match the higher-order $\calO(n^{-2})$ standard error rate for RQMC-lat. Using higher-order smoothness SI kernels from \Cref{def:si_kernel} may help to match convergence rates. We also found BQMC-lat often required a burn-in period of around $2^9$ after which the standard error significantly dropped and settled into its asymptotic convergence rate. This was not the case for BQMC-net which yielded steady standard error convergence across all tested sample sizes. 

While the multilevel BQMC-lat methods showed performance on par with the RQMC-lat methods in terms of true error, their standard errors often proved unreliable. In the Asian option and Darcy flow problems, the BQMC-lat standard errors have rate constants which are at least an order of magnitude worse than their true errors for moderate to large budgets. In fact, for the Darcy flow problem, the lattice standard error is worse than that of IID Monte Carlo. We suspect the inaccurate BQMC-lat standard errors are again due to their unreasonable periodicity assumption. That said, we did observe accurate BQMC-lat standard error estimates for the Lookback option and elliptic PDE examples. 

\end{document}